\documentclass{aa}
\usepackage{txfonts}
\usepackage[authoryear]{natbib}
\usepackage{natbib}
\usepackage{amssymb}
\bibpunct{(}{)}{;}{a}{}{,}
\usepackage{epsf}
\usepackage{graphicx}
\usepackage{epsfig}
\usepackage{graphics}
\usepackage{subfigure}
\usepackage[english]{babel}
\usepackage{rotating}
\usepackage{color}

\newcommand{\teff}{$T_{\rm{eff}}$}
\newcommand{\logg}{$\log g$}
\newcommand{\lL}{\ifmmode \log \frac{L}{L_{\sun}} \else $\log \frac{L}{L_{\sun
}}$\fi}

\newcommand{\vsini}{$V$~sin$i$}

\newcommand{\vmac}{$v_{\rm mac}$}

\newcommand{\kms}{km~s$^{-1}$}
\newcommand{\msun}{M$_{\sun}$}

\begin{document}

\title{The MiMeS Survey of Magnetism in Massive Stars: \\
CNO surface abundances of Galactic O stars\thanks{Based on observations obtained at 1) the Telescope
Bernard Lyot (USR5026) operated by the Observatoire Midi-Pyr\'en\'ees, Universit\'e de Toulouse (Paul Sabatier), Centre National de la Recherche Scientifique of France; 2) at the Canada-France-Hawaii Telescope (CFHT) which is operated by the National Research Council (NRC) of Canada, the Institut National des Science de l'Univers of the Centre National de la Recherche Scientifique (CNRS) of France, and the University of Hawaii; 3) at the ESO/La Silla Observatory under program ID 187.D-0917.}}
\author{F. Martins\inst{1}
\and A. Herv\'e\inst{1}
\and J.-C. Bouret\inst{2}
\and W. Marcolino\inst{3}
\and G.A. Wade\inst{4}
\and C. Neiner\inst{5}
\and E. Alecian\inst{6,5}
\and J. Grunhut\inst{7}
\and V. Petit\inst{8}
\and the MiMeS collaboration
}
\institute{LUPM, Universit\'e Montpellier 2, CNRS, Place Eug\`ene Bataillon, F-34095 Montpellier, France  \\
           \email{fabrice.martins@univ-montp2.fr}
\and
LAM--UMR 6110, CNRS \& Universit\'e de Provence, rue Fr\'ed\'eric Joliot-Curie, F-13388, Marseille Cedex 13, France\\
\and
Observatorio do Valongo, Universidade Federal do Rio de Janeiro, Ladeira Pedro Ant\'onio, 43, CEP 20080-090, Brasil\\
\and
Dept. of Physics, Royal Military College of Canada, PO Box 17000, Stn Forces, Kingston, Ontario K7K 7B4, Canada \\
\and
LESIA, Observatoire de Paris, CNRS UMR 8109, UPMC, Universit\'e Paris Diderot, 5 place Jules Janssen, 92190, Meudon, France\\
\and
UJF-Grenoble 1/CNRS-INSU, Institut de Plan\'etologie et d'Astrophysique de Grenoble, UMR 5274, 38041, Grenoble, France \\
\and
ESO, Karl-Schwarzschild-Str. 2, D-85748 Garching, Germany \\
\and
Dept. of Physics \& Astronomy, University of Delaware, Newark, DE, USA
}

\date{Received / Accepted }

\abstract
{The evolution of massive stars is still partly unconstrained. Mass, metallicity, mass loss and rotation are the main drivers of stellar evolution. Binarity and magnetic field may also significantly affect the fate of massive stars.}
{Our goal is to investigate the evolution of single O stars in the Galaxy. }
{For that, we use a sample of 74 objects comprising all luminosity classes and spectral types from O4 to O9.7. We rely on optical spectroscopy obtained in the context of the MiMeS survey of massive stars. We perform spectral modelling with the code CMFGEN. We determine the surface properties of the sample stars, with special emphasis on abundances of carbon, nitrogen and oxygen.}
{Most of our sample stars have initial masses in the range 20 to 50 \msun. We show that nitrogen is more enriched and carbon/oxygen more depleted in supergiants than in dwarfs, with giants showing intermediate degrees of mixing. CNO abundances are observed in the range of values predicted by nucleosynthesis through the CNO cycle. More massive stars, within a given luminosity class, appear to be more chemically enriched than lower mass stars. We compare our results with predictions of three types of evolutionary models and show that, for two sets of models, 80\% of our sample can be explained by stellar evolution including rotation. The effect of magnetism on surface abundances is unconstrained.}
{Our study indicates that, in the 20-50 \msun\ mass range, the surface chemical abundances of most single O stars in the Galaxy are fairly well accounted for by stellar evolution of rotating stars.}


\authorrunning{Martins et al.}
\titlerunning{CNO abundances of Galactic O stars}

\maketitle

%%%%%%%%%%%%%%%%%%%%%%%%%%%%%%%%%%%%%%%%%%%%%%%%%%%%%%%%%%%%%%%%%%%%%%%%%%%%%%%%%%%%%%%%%%%%%%%%%%%%%%%%%%%%%%%%%%%%%%%%%%%%%%%
%%%%%%%%%%%%%%%%%%%%%%%%%%%%%%%%%%%%%%%%%%%%%%%%%%%%%%%%%%%%%%%%%%%%%%%%%%%%%%%%%%%%%%%%%%%%%%%%%%%%%%%%%%%%%%%%%%%%%%%%%%%%%%%
\section{Introduction}
\label{s_intro}

Massive stars are usually defined as stars with initial masses larger than about 8 \msun. They have short lives (about 5 to 20 Myr) and explode as core-collapse supernovae. Before ending their life, they change appearance while crossing various phases of their evolution. Born as O and early B stars, they become blue supergiants. The most massive ones may then enter the unstable phase of Luminous Blue Variables (LBV), while the others evolve into red supergiants before their final explosion. Most stars with initial masses higher than 25 \msun\ evolve back to the hot part of the Hertzsprung-Russell diagram (HRD) due to the action of strong stellar winds which peel off their outer layers: they appear as Wolf-Rayet (WR) stars of different flavor (WN, WC or WO) depending on the strength of mass loss and the advancement of nucleosynthesis in their internal layers. 

This general picture \citep[the Conti scenario,][]{conti75} has been refined over the years \citep[e.g.][]{paul07} but still suffers from many uncertainties. For instance, the exact role of the LBV phase is a matter of debate \citep{smith06}. The nature of type Ib/Ic supernovae is not clear \citep[e.g.][]{groh13}. The oxygen-rich WR stars (type WO), once thought to be more evolved than any other WR stars, may simply be the hottest WC stars \citep{tramper13}. Mass is also not the only parameter governing the evolution of massive stars which, especially in the advanced phases, depends on mass loss. The strength of radiatively driven winds scales with metallicity \citep{vink01,paul02,mokiem07}. Consequently, the evolution of massive stars is affected by their metal content. Another major ingredient of massive star evolution is rotation. Through its mechanical effects, it flattens stars which in turn modifies the surface properties (local temperature and gravity). This results in asymmetrical stellar winds, which affects angular momentum loss as well as the strength of mass loss itself \citep{mm00}. In addition, rotation triggers hydrodynamical instabilities in the internal layers, leading to transport of angular momentum and mixing \citep{mm96,hl00}. Material produced in the core or nuclear burning shells is transported to the surface, changing the appearance of stars. The presence of a magnetic field may affect mixing triggered by rotation, and subsequently impact the evolution \citep{mm05b,petro05}. Last but not least, the presence of a companion may drastically change the evolution of massive stars \citep{langer12}: tidal interaction and mass transfer can severely modify the rotation, surface chemical composition and mass of binary components \citep{petro05,dm09,song13}.

Surface abundances are a key to the understanding of single star evolution. The more evolved a star is, the heavier the elements detected on its surface. These elements are seen because mixing transports them to the surface and, in the advanced phases, because mass loss removes external layers, pushing the stellar surface deeper where elements were produced by nuclear burning. In OB stars, the former effect (mixing) dominates. Surface abundances are thus a direct signature of nucleosynthesis and rotation. Since the main nuclear reactions in OB stars are those of the CNO cycle, helium, carbon, nitrogen and oxygen are the key elements probing internal mixing during early evolution. Evolutionary calculations show that during the main sequence, the surface nitrogen abundance increases, while carbon and oxygen are depleted \citep{mm00,langer12}. At the same time, the surface helium content increases too, although by only a smaller fraction than C, N and O because He is already the second most abundant element on the surface. Evolutionary calculations also predict that the degree of chemical mixing depends on metallicity and initial mass, mainly because of differences in the internal rotational velocity profile \citep[e.g.][]{maeder14}. \citet{aerts14} pointed out that pulsations may be important too.

Tests of the predictions of evolutionary models including the effects of rotation have been mainly performed on B stars. The VLT Flames survey of massive stars \citep{evans05,evans06} collected optical spectra of hundreds of OB stars in the Milky Way, the Large Magellanic Cloud (LMC) and the Small Magellanic Cloud (SMC). Stellar and wind properties were derived by \citet{mokiem06,mokiem07}, \citet{hunter07} and \citet{trundle07}. \citet{hunter08} compared the surface nitrogen content with projected rotational velocities (\vsini) for LMC B stars. They found that about 60\% of their sample was well accounted for by evolutionary calculations. The remaining 40\% were either too N-rich or too N-poor for their \vsini. In particular, a group of stars with low \vsini\ showed significant nitrogen enrichment. This study relied on stars with masses between 10 and 20 \msun\ which may be too wide to separate the effects of rotation and mass on chemical mixing. To better account for mass dependence of chemical enrichment, \citet{brott11b} performed population synthesis calculations and confirmed the results of \citet{hunter08}. However, \citet{maeder09} concluded that only 20\% of B stars did not follow the predictions of evolutionary models. \citet{hunter09} extended the work of \citet{hunter08} to the SMC and the Milky Way. They confirmed the presence of N-rich B stars with relatively low \vsini\ that cannot be explained by standard evolution, unless, in the case of supergiants, they are post-red supergiant objects.

\citet{przy10}, building on previous works \citep{np06,np07,przy06}, used not only nitrogen but also carbon and oxygen to investigate the evolution of surface chemistry of B stars. They showed that the ratios N/C and N/O were tightly correlated, as expected from the physics of CNO burning. In addition, they showed that evolutionary models could qualitatively account for the range of observed N/C ratios in most cases, although the models seemed to slightly underpredict the amplitude of chemical mixing. \citet{maeder14} re-analyzed part of the VLT Flames survey B stars sample in light of these results. They again found a very tight correlation between N/C and N/O. They compared this trend with different evolutionary models but could not favour one over the other because of the too large uncertainties in the observed abundance ratios. The general conclusion of these studies is that evolutionary models including rotation reproduce well the surface chemical properties of most B stars, but an uncertain fraction of them may require additional physics. So far, a given set of models for B stars cannot be preferred. 

Chemical abundance determinations of more massive O stars are less numerous. Although mixing is expected to be stronger in more massive stars, and thus easier to test, it is also more difficult to determine CNO abundances in O stars. Compared to B stars, non-LTE effects are much stronger and line formation is harder to reproduce in atmosphere models \citep{rg11,mh12}. \citet{rg12} determined nitrogen abundances of LMC O stars. Their sample was limited to 20 objects, but they seemed to find a large population of N-rich slow rotators as is found among B stars. \citet{jc13} obtained CNO abundances of 23 SMC O dwarfs. Based on the N/C and N/O ratios, they found that only 9\% of the stars were clearly not compatible with evolutionary models (another 26\% being only marginally compatible). In the Galaxy, \citet{jc12} provided abundances of 8 supergiants. Four objects were correctly explained by models with rotation, while four showed too large N/C ratios. \citet{martins12} analyzed 8 Galactic dwarfs and concluded that their nitrogen surface content was consistent with evolutionary tracks of appropriate initial mass. Clearly, our current understanding of surface abundances of O stars is incomplete. Samples are so far limited to about 10-20 objects of a given luminosity class, at a given metallicity. 

In this paper, we present a significant improvement of this situation. We determine the stellar parameters and CNO surface abundances of 74 Galactic O stars. We compare the N/C and N/O values with predictions of various grids of evolutionary models. We investigate the effects of mass and metallicity on chemical mixing. The paper is organized as follows: the sample and the observations are described in Sect.\ \ref{s_obs}; the analysis method and results are presented in Sect.\ \ref{s_mod}; the CNO surface abundances are discussed in Sect.\ \ref{s_ab} and the conclusions are given in Sect.\ \ref{s_conc}.

%%%%%%%%%%%%%%%%%%%%%%%%%%%%%%%%%%%%%%%%%%%%%%%%%%%%%%%%%%%%%%%%%%%%%%%%%%%%%%%%%%%%%%%%%%%%%%%%%%%%%%%%%%%%%%%%%%%%%%%%%%%%%%%
%%%%%%%%%%%%%%%%%%%%%%%%%%%%%%%%%%%%%%%%%%%%%%%%%%%%%%%%%%%%%%%%%%%%%%%%%%%%%%%%%%%%%%%%%%%%%%%%%%%%%%%%%%%%%%%%%%%%%%%%%%%%%%%
\section{Observations and sample}
\label{s_obs}

The observations were performed within the MiMeS survey of massive stars (Wade et al.\, in prep.). The survey was designed to investigate the magnetic properties of massive OB stars. 111 O stars were observed, including nine magnetic stars. The criteria were the visibility from one of the three sites of observation (see below), the V-band apparent magnitude so that the high signal-to-noise ratio required for spectropolarimetric observations can be reached (4$<$V$<$8), and (in most cases) the existence of UV spectroscopy (mainly from \textit{IUE}). Large observing programs were granted for the observations. The three sites and instruments are:

\begin{itemize}

\item \textit{Canada France Hawaii Telescope}: the spectropolarimeter ESPaDOnS was used. It is an \'echelle spectrograph with a resolving power of 65000 equipped with polarimetric capabilities. The wavelength coverage is 3700 \AA\ to 1.05 $\mu$m. The faintest northern targets were observed with ESPaDOnS.

\item \textit{Pic du Midi Observatory}: NARVAL, a twin of ESPaDOnS, is mounted on the 2 meters T\'elescope Bernard Lyot. The wavelength coverage and spectral resolution are the same as ESPaDOnS. The brightest northern targets were observed with NARVAL.

\item \textit{La Silla Observatory}: the HARPS spectrograph equipped with its polarimetric module (HARPSpol) mounted on the ESO 3.6m telescope was used to observe southern targets. The spectral resolution of HARPSpol is 105000 and the wavelength coverage is 3800-7000 \AA. 

\end{itemize}

\noindent For each star, at least one sequence of four polarimetric observations (to obtain Stokes V profiles) was performed. Detecting Zeeman signatures requires signal-to-noise ratios (SNR) of several hundreds (the exact value depending on the strength of the magnetic field, the spectral type and the rotational velocity). The distribution of SNR was quite broad, extending from below 100 up to more than 2000. The median SNR of all stars was 800 per pixel. This makes the resulting spectra perfectly suited for abundance analysis, since even the weakest metallic lines are resolved and detected. The data were reduced using the \textit{Libre Esprit} package. A full description of the reduction process is provided in e.g. \citet{wade11}. The data are described in greater detail by Wade et al.\ (in prep). 

Our goal is to investigate the chemical properties of single O stars. We began by renormalising all observations. Out of the 111 objects observed by MiMeS, we then excluded all the known spectroscopic binaries (SB1 and SB2). This includes stars with known orbital parameters as well as stars for which only hints of binarity have been reported (radial velocity variations, composite spectrum). We ended up with a sample of 67 O stars. The distribution with respect to luminosity classes is the following: 23 dwarfs, 4 subgiants, 16 giants, 6 bright giants and 18 supergiants. The spectral type distribution is: 45 stars with O9.7 $<$ ST $<$ O8, 17 stars with O7.5 $<$ ST $<$ O6 and 5 stars with ST $<$ O5.5. Thus the sample contains about 2/3 of late type O stars.  We have also included seven O stars with a magnetic field: $\theta^1$Ori~C, HD~108, HD~57682, HD~148937, HD~191612, Tr16-22\footnote{reported to be magnetic by \citet{naze12}. The spectrum used in the present paper was obtained from the ESO/FEROS archive.} and CPD-28 2561. The magnetic stars HD~47129, HD~37742 and NGC1624-2 were not included: the first two objects are binaries while the second has such a strong field that lines are affected by significant Zeeman broadening, rendering the spectroscopic analysis uncertain.

%%%%%%%%%%%%%%%%%%%%%%%%%%%%%%%%%%%%%%%%%%%%%%%%%%%%%%%%%%%%%%%%%%%%%%%%%%%%%%%%%%%%%%%%%%%%%%%%%%%%%%%%%%%%%%%%%%%%%%%%%%%%%%%
%%%%%%%%%%%%%%%%%%%%%%%%%%%%%%%%%%%%%%%%%%%%%%%%%%%%%%%%%%%%%%%%%%%%%%%%%%%%%%%%%%%%%%%%%%%%%%%%%%%%%%%%%%%%%%%%%%%%%%%%%%%%%%%
\section{Atmosphere models and spectroscopic analysis}
\label{s_mod}

We have used the code CMFGEN \citep{hm98} to determine the fundamental properties of the sample stars. CMFGEN computes non-LTE and spherical models of massive stars atmospheres. It includes winds and line-blanketing. The hydrodynamical structure (density, velocity) is given as an input. The velocity structure is constructed from a combination of an inner structure and a $\beta$ velocity law in the outer part. Once the level populations have converged (see below) the radiative acceleration is calculated and a new inner structure is computed and connected to the same $\beta$ velocity law. Two of these global iterations have been performed in our computation, resulting in a self-consistent hydrodynamical structure below the connection point. The density structure is obtained from the mass conservation equation. The level populations are calculated through the rate equations. A super-level approach is used to reduce the size of the problem (and thus the computing time). We have included the following elements in our calculations: H, He, C, N, O, Ne, Si, S, Ar, Ca, Fe and Ni. 
Once the atmospheric structure is obtained, a formal solution of the radiative transfer equation, including the proper line profiles, is performed. A depth variable microturbulent velocity starting from 10 \kms\ at the photosphere and reaching 10\%\ of the terminal velocity at the top of the atmosphere is assumed. The resulting spectrum is convolved with the appropriate rotational and macroturbulent velocities (see below) and is compared to the observed spectrum\footnote{Instrumental broadening being of the order of 4 \kms, it is negligible}.

\begin{figure}[t]
\centering
\includegraphics[width=9cm]{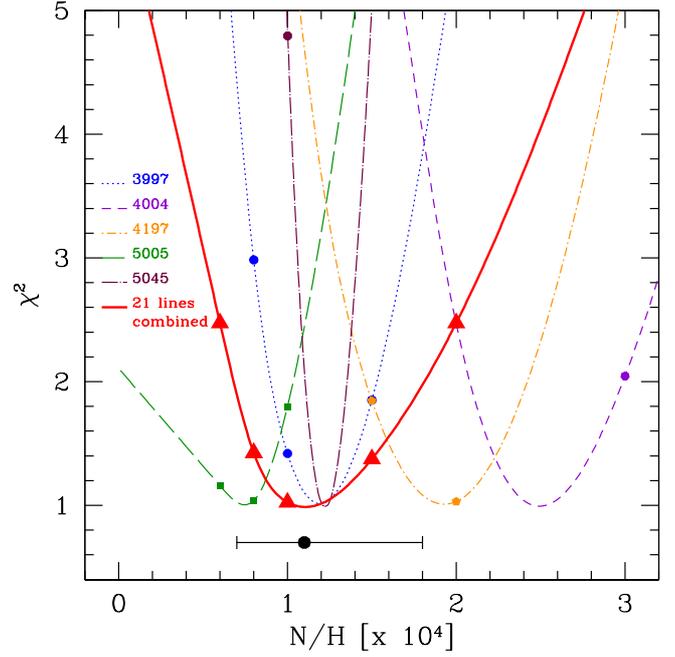}
\caption{Illustration of the effect of the choice of the spectral lines for the determination of the nitrogen abundance of star HD~207538. The thin broken lines shows the normalized $\chi^2$ curve when the fit is performed on individual lines (wavelength indicated in the figure). The bold solid curve shows the $\chi^2$ of the analysis combining 21 \ion{N}{ii} and \ion{N}{iii} lines. The black dot shows the preferred value of N/H resulting from the fit of the combined lines, together with the error bars. }
\label{check_N}
\end{figure}

We have relied on the following diagnostics to constrain the fundamental parameters:

\begin{itemize}

\item \textit{Rotational velocity}: \vsini\ is obtained by the Fourier transform method \citep{gray76,sergio07}. We have used \ion{O}{iii}~5592 as our main indicator. In case the spectrum in that region was not of good enough quality, we relied on \ion{C}{iv}~5802 or \ion{He}{i}~4713. The synthetic spectra computed by CMFGEN have been convolved by a rotational profile corresponding to the position of the first zero in the Fourier transform. The uncertainty on \vsini\ is $\sim$10 \kms.

\item \textit{Macroturbulent velocity}: We fitted the \ion{O}{iii}~5592 line with different synthetic spectra convolved by a Gaussian profile mimicking macroturbulence (in addition to the convolution by rotational broadening). The Gaussian profile\footnote{Note that this definition is slightly different from that commonly used in studies of macroturbulence where the profile is $\propto e^{-\frac{v^2}{v_{mac}^2}}$.} was of the form $\frac{1}{v_{mac}\sqrt{2\pi}} e^{-\frac{v^2}{2 v_{mac}^2}}$. The uncertainty on these measurements is $\sim$10 \kms.

\item \textit{Effective temperature}: the ionization balance method based on helium lines was used. In practice, the following lines were selected: \ion{He}{i}~4026, \ion{He}{i}~4388, \ion{He}{i}~4471, \ion{He}{i}~4713, \ion{He}{i}~4922, \ion{He}{ii}~4200, \ion{He}{ii}~4542, \ion{He}{ii}~5412. We have excluded \ion{He}{i}~5876 and \ion{He}{ii}~4686 since they are usually affected by winds. Given the good quality of the data and the relatively large number of diagnostic lines, the typical uncertainty on our determinations is 1000 K.

\item \textit{Surface gravity}: the wings of Balmer lines are the main diagnostic of \logg. Due to the \'echelle nature of our spectra, the normalization around Balmer lines was difficult. We often observed that a value of \logg\ leading to a good fit to one line was not necessarily the best value for other lines. Consequently, we estimate an uncertainty of about 0.15 dex on \logg\ as representative of our determinations.

\end{itemize}

\noindent In absence of strong constraint on the distance of most of the stars, we decided to adopt the luminosities. We used the calibrations of \citet{msh05} for that purpose. The wind parameters (mass loss rate, terminal velocity, clumping) were set from both H$\alpha$ and the UV lines \ion{Si}{iv}~1400, \ion{C}{iv}~1550, \ion{He}{ii}~1640 and \ion{N}{iv}~1720\footnote{We used spectra from the \textit{IUE} archive.}. However, since our prime focus was to constrain the surface stellar parameters and abundances, we did not push the analysis of the wind parameters as far as that of the surface parameters. We simply ensured that they lead to a reasonable fit of the above lines. 

In addition to the classical parameters listed above, we took special care in determining the surface abundances of CNO elements. Once the fundamental parameters were constrained, we ran several models changing only the CNO abundances. We then performed a $\chi^{2}$ analysis to estimate the abundances (and the associated uncertainties) giving the best fit to selected lines. For each element, the $\chi^{2}$ function was renormalized so that the minimum has a value of 1.0. The 1$\sigma$ uncertainty was set to the abundances for which $\chi^{2}=2.0$. This process is illustrated in Fig.\ \ref{check_N} for the case of carbon in HD~207538. The choice of the diagnostic lines depends on the quality of the spectrum and on the spectral type. Here is the list of the lines among which we made the selection of the diagnostics used in the $\chi^{2}$ analysis:

\begin{itemize}

\item \textit{Carbon}: \ion{C}{iii}~4068-70, \ion{C}{iii}~4153, \ion{C}{iii}~4156, \ion{C}{iii}~4163, \ion{C}{iii}~4187,  \ion{C}{ii}~4267, \ion{C}{iii}~4325, \ion{C}{iii}~4666, \ion{C}{iii}~5246, \ion{C}{iii}~5353, \ion{C}{iii}~5272, \ion{C}{iii}~5826, \ion{C}{iii}~6205, \ion{C}{iii}~6744. One to 14 lines were used depending on the star and quality of the spectrum. We excluded  \ion{C}{iii}~4647-50-51 and \ion{C}{iii}~5696 since their formation process depends on fine details of atomic physics and of the modelling \citep{mh12}.

\item \textit{Nitrogen}: \ion{N}{ii}~3995, \ion{N}{ii}~4004, \ion{N}{ii}~4035, \ion{N}{ii}~4041, \ion{N}{iii}~4044,  \ion{N}{iii}~4196, \ion{N}{ii}~4447, \ion{N}{iii}~4511,  \ion{N}{iii}~4515, \ion{N}{iii}~4518, \ion{N}{iii}~4524, \ion{N}{ii}~4607, \ion{N}{ii}~4803, \ion{N}{iii}~4907, \ion{N}{ii}~5001, \ion{N}{ii}~5005, \ion{N}{ii}~5011, \ion{N}{ii}~5026, \ion{N}{ii}~5045, \ion{N}{iv}~5200, \ion{N}{iv}~5204, \ion{N}{ii}~5676, \ion{N}{ii}~5680. Four to 22 lines were used.

\item \textit{Oxygen}:  \ion{O}{ii}~3792, \ion{O}{ii}~3913, \ion{O}{ii}~3955, \ion{O}{ii}~3963, \ion{O}{ii}~4277-78,  \ion{O}{ii}~4284, \ion{O}{ii}~4305, \ion{O}{ii}~4318, \ion{O}{ii}~4321, \ion{O}{ii}~4368, \ion{O}{ii}~4416, \ion{O}{ii}~4418, \ion{O}{ii}~4592, \ion{O}{ii}~4597, \ion{O}{ii}~4603, \ion{O}{ii}~4611, \ion{O}{ii}~4663, \ion{O}{ii}~4678, \ion{O}{ii}~4700, \ion{O}{ii}~4707,  \ion{O}{iii}~5592. Two to 20 lines were used.

\end{itemize}

\noindent Figure \ref{check_N} illustrates the importance of selecting as many lines as possible for the abundance determinations. Each line used individually gives a different abundance. Taking into account as many lines as possible gives an average value of N/H and provides a better estimate of the uncertainty. The errors are due to uncertainties in atomic data, in the line formation processes (non-LTE effects) and to the accuracy of the ionization balance in our models for the estimated effective temperature (especially when lines of consecutive ions from the same element are used). The final errors range from 20\% to 100\% in the worst cases. They are larger than errors quoted for B stars \citep{przy10,nieva14,maeder14} for the reasons mentioned above (non-LTE effects are much more severe in O stars). Our error determination does not include uncertainties on \teff\ and/or \logg. In a previous study \citep{Omag12}, we used a different approach: for different values of \teff\ and \logg\ we ran models with various abundances and selected, for each set of temperatures and gravities, the best fit abundance. We then computed the average and standard deviation of these measurements that we adopted as the derived abundance and uncertainty. We checked on one example that both strategies \citep[the present one and that of][]{Omag12} lead to similar results.

\begin{figure*}[t]
\centering
\includegraphics[width=18cm]{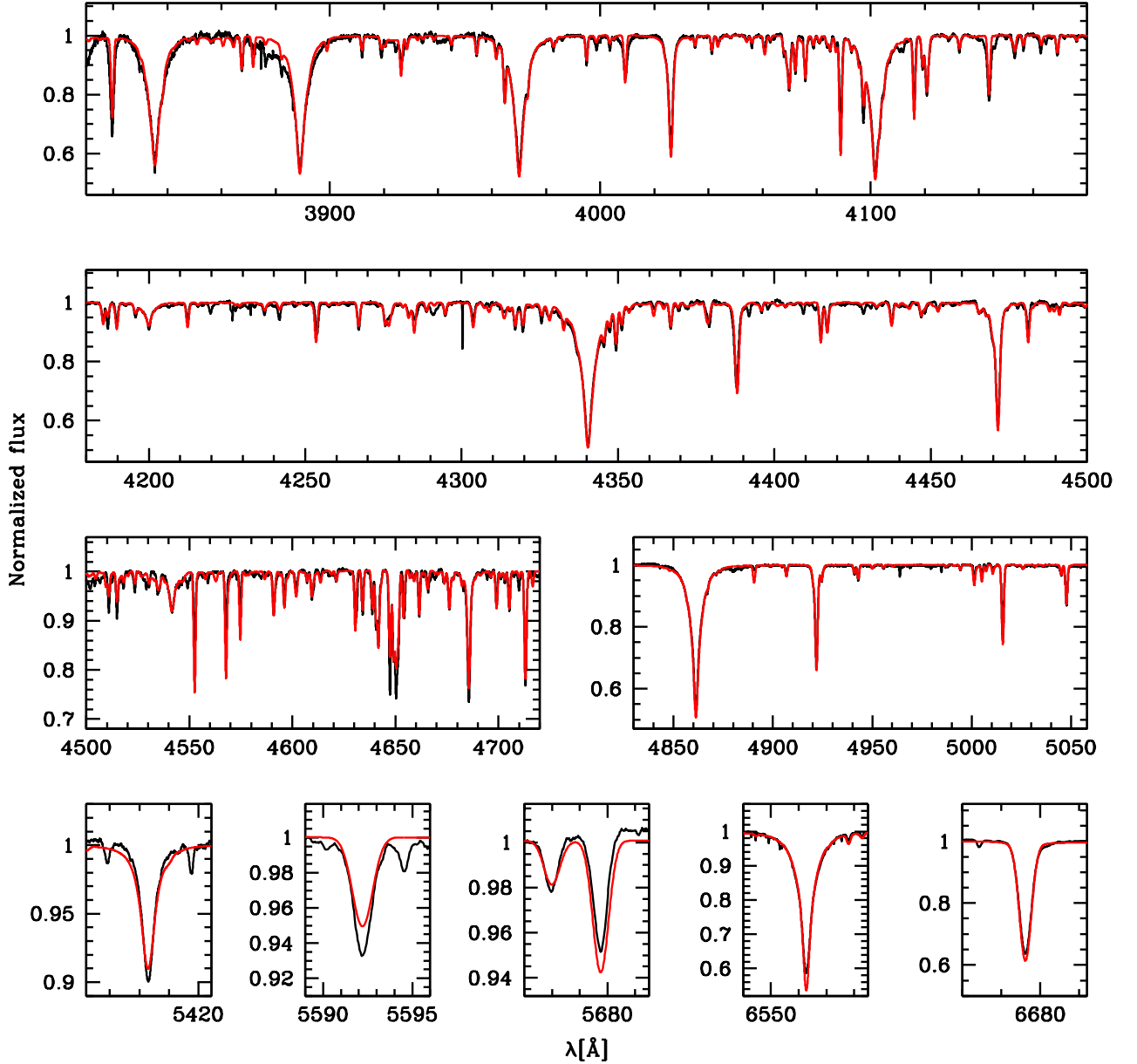}
\caption{Best fit model (red) of the observed spectrum (black) of HD~207538. }
\label{fit_207538}
\end{figure*}

Fig.\ \ref{fit_207538} shows a typical fit we obtain, with HD~207538 as an example. The quality is usually very good. Some problems, mostly due to normalization of the \'echelle spectra, remain in a few regions (e.g. the blue wing of H$\epsilon$). For the reasons given previously, the fit of \ion{O}{iii}~5592 is not perfect: fitting this line would require a larger O/H ratio, but at the cost of degrading the fit of many other oxygen lines, especially the numerous \ion{O}{ii} lines between 4590 and 4615 \AA.

\begin{figure}[t]
\centering
\includegraphics[width=9cm]{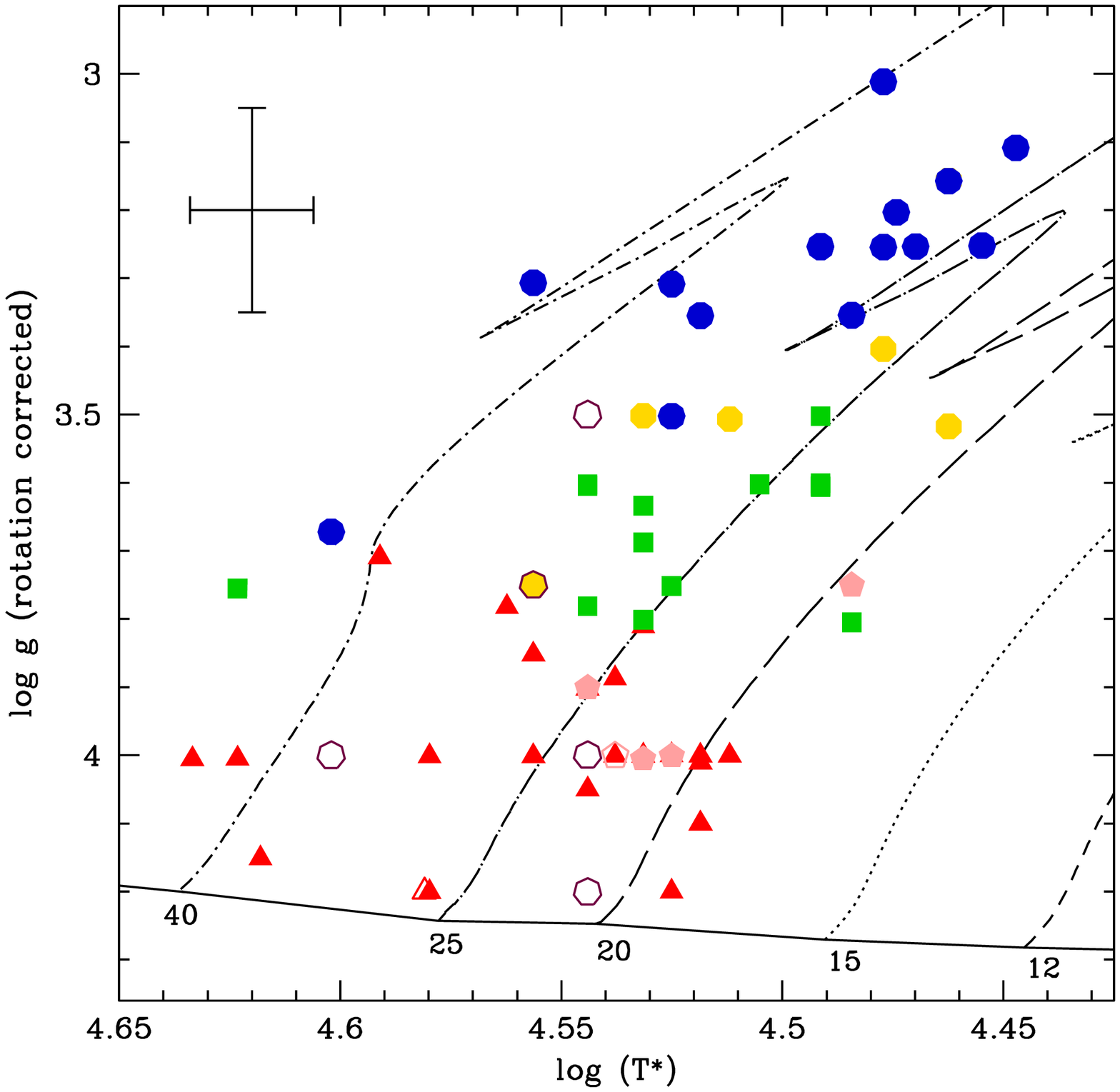}
\caption{\logg\ - \teff\ diagram of the sample stars. Red triangles are dwarfs; pink pentagons are subgiants; green squares are giants; yellow octagons are bright giants; blue circles are supergiants. Open symbols are magnetic stars, with magenta heptagons being Of?p stars. Evolutionary tracks including rotation are from \citet{ek12}. Typical error bars are indicated in the figure.}
\label{fig_hr}
\end{figure}

Fig.\ \ref{fig_hr} shows the position of the sample stars in the \logg\ -- \teff\ diagram. In absence of accurate constraints on their distance and thus on their luminosity, this diagram better accounts for their evolutionary status. Tracks including rotation from \citet{ek12} are overplotted. Some stars have similar \teff\ and \logg\, explaining that the number of apparent points in Fig.\ \ref{fig_hr} is smaller than 74. Most of the sample stars have initial masses between 20 and 50 \msun. Dwarfs, giants and supergiants are clearly separated in \logg. Subgiants and bright giants are nested respectively between dwarfs and giants and giants and supergiants, as expected. Dwarfs have surface gravities between 3.8 and 4.2. This range is the same across the  20--50 \msun\ mass range. Giants have \logg$\approx$3.7 in the highest masses probed and \logg$\approx$3.5 in the lowest mass bin. Supergiants have about the same \logg\ as giants around 40 \msun, but \logg$\approx$3.2 at M=20 \msun. These trends are quantitatively consistent with those determined by \citet{msh05}. 

The location of the supergiants in Fig.\ \ref{fig_hr} corresponds to the end of the main sequence in the evolutionary tracks of \citet{ek12}. As noted by \citet{mp13}, this is an indication that the size of the convective core of the Ekstr$\ddot{o}$m et al. models is appropriate, at least in the 20-50 \msun\ mass bin we probe here. The magnetic stars cover a range of surface gravities. HD~191612 has the same gravity and effective temperature as the bright giant HD~34656.

\begin{table*}
\begin{center}
\caption{Properties of the sample stars.} \label{tab_param}
\begin{tabular}{lccccccccc}
\hline
Star        & ST            &   Teff     &   logg    &  \vsini\  &  \vmac  &  C/H         &    N/H      & O/H   \\    
            &               & [kK]       &           &  [\kms]   &  [\kms] &  [10$^{-4}$]  &  [10$^{-4}$] &  [10$^{-4}$] \\
\hline
HD~13745    & O9.7II(n)     & 29          &  3.1     &  175 &   49    &  1.0$^{+0.2}_{-0.2}$ &  2.0$^{+1.6}_{-0.8}$ & 4.1$^{+2.5}_{-2.5}$  \\
HD~14633    & ON8.5V        & 34          &  3.8     &  100 &   76    &  $<$0.3     &  5.2$^{+5.0}_{-2.5}$ & 1.7$^{+0.9}_{-0.8}$  \\
HD~24431    & O9III         & 33.5       &  3.75     &  37  &   43    &  1.3$^{+0.3}_{-0.3}$ &  0.8$^{+0.4}_{-0.4}$ & 3.9$^{+2.3}_{-1.9}$  \\
HD~24912    & O7.5III(n)((f)) & 34       &  3.60     &  180 &   50    &  1.3$^{+0.5}_{-0.5}$ &  4.2$^{+1.3}_{-1.1}$ & 3.8$^{+1.6}_{-1.0}$ \\
HD~30614    & O9Ia          & 29.5       &  3.25     &  50  &   47    &  0.6$^{+0.4}_{-0.3}$ &  3.0$^{+1.8}_{-1.6}$ & 2.7$^{+2.2}_{-2.0}$ \\
HD~34078    & O9.5V         & 33         &  4.0      &  25  &   5     &  2.6$^{+1.0}_{-1.0}$ &  0.6$^{+0.3}_{-0.3}$  & 5.5$^{+2.5}_{-1.9}$   \\  
HD~34656    & O7.5II(f)     & 36         &  3.75     &  30  &   40    &  0.7$^{+0.2}_{-0.2}$ &  4.8$^{+2.5}_{-1.8}$  & 1.7$^{+1.8}_{-0.6}$  \\
HD~35619    & O7.5V((f))    & 35         &  3.90     &  56  &   30    &  1.3$^{+0.4}_{-0.4}$ & $<$0.8 & 3.3$^{+3.0}_{-1.8}$ \\
HD~36512    & O9.7V         & 32.5      & 4.0        &  20  &   10    &  2.4$^{+0.8}_{-0.8}$ & 0.6$^{+0.2}_{-0.2}$ & 5.2$^{+1.9}_{-1.9}$ \\
HD~36861    & O8III((f))    & 35          &  3.75    &  38  &   38    &  2.2$^{+0.5}_{-0.4}$ &  1.5$^{+0.4}_{-0.3}$ & 4.7$^{+2.0}_{-1.4}$  \\
HD~36879    & O7V(n)((f))z  & 36.5        &  3.75    &  200 &   --    &  1.6$^{+0.4}_{-0.4}$ &  2.5$^{+0.9}_{-0.4}$ & 2.1$^{+1.2}_{-0.9}$  \\
HD~38666    & O9.5V         & 33         & 4.0       & 111  & 16      &  2.1$^{+0.5}_{-0.5}$ & $<$0.6 & 2.7$^{+1.7}_{-1.0}$ \\
HD~42088    & O6V           & 38        &  4.0       &  41  & 37      &  2.0$^{+0.3}_{-0.3}$ & $<$1.0 & 2.9$^{+1.7}_{-0.9}$ \\
HD~46056    & O8Vn          & 34.5      & 3.75       & 330  & --      &  1.9$^{+0.3}_{-0.3}$ & 0.6$^{+0.2}_{-0.2}$ & 2.8$^{+1.8}_{-1.8}$ \\
HD~46106    & O9.7III(n)    & 30.5      & 3.8        &  79  & 62      &  0.7$^{+0.5}_{-0.5}$ & $<$0.6 & 1.1$^{+0.5}_{-0.5}$ \\
HD~46150    & O5V((f))z     & 42          &  4.0     &  100 &   38    & 2.3$^{+0.5}_{-0.5}$  & 5.4$^{+1.9}_{-1.6}$  & 2.0$^{+0.3}_{-0.3}$  \\
HD~46202    & O9.2V         & 33.5       &  4.2      &  15  &   13    & 1.9$^{+0.4}_{-0.4}$ &  0.95$^{+0.3}_{-0.3}$  & 3.4$^{+1.5}_{-1.3}$ \\
HD~46223    & O4V((f))      & 43          & 4.0      & 100  & 32      &  0.7$^{+0.3}_{-0.3}$ & 9.0$^{+5.5}_{-4.2}$  & 2.4$^{+0.6}_{-0.5}$  \\
HD~46485    & O7V((f))nz    & 36       &  3.75       &  300 &   --    & 2.7$^{+0.7}_{-0.5}$  & 0.9$^{+0.2}_{-0.2}$ & 4.4$^{+2.2}_{-2.0}$ \\
HD~46966    & O8.5IV        & 35         &  3.9      &  33  &   36    & 1.1$^{+0.5}_{-0.5}$ & 1.0$^{+0.6}_{-0.4}$ & 2.9$^{+2.2}_{-1.1}$  \\
HD~47432    & O9.7Ib        & 29         &  3.15     &  50  &   55    &  $<$0.8     &  2.0$^{+2.1}_{-1.1}$ & 2.7$^{+2.7}_{-2.0}$ \\
HD~55879    & O9.7III       & 31          &  3.6     &  25  &   25    &  0.7$^{+0.4}_{-0.3}$ &  1.1$^{+0.5}_{-0.4}$ & 3.0$^{+1.0}_{-1.0}$  \\
HD~66788    & O8V           & 34.5         &  4.0      &  24  & 31      &  1.0$^{+0.5}_{-0.5}$ & $<$1.0  & 4.3$^{+3.0}_{-1.8}$   \\
HD~66811\tablefootmark{a}   & O4I(n)fp &  40 & 3.64  &  210 & 90      &  0.04$^{+0.02}_{-0.02}$ & 12.6$^{+4.9}_{-4.9}$  & 1.35$^{+0.9}_{-0.9}$   \\
HD~69106    & O9.7IIn       &  29  & 3.4             &  320 & --      &  0.4$^{+0.2}_{-0.1}$ &  $<$1.0 & 2.5$^{+1.2}_{-0.8}$  \\
HD~93250    & O4III(fc)     &  42     & 3.75         &  90  &  52     & 2.1$^{+0.6}_{-0.6}$ &  3.5$^{+1.7}_{-1.2}$ & 2.0$^{+0.4}_{-0.4}$  \\
HD~149038   & O9.7Iab       &  28.5    &  3.25       &  38  &   40    &  0.5$^{+0.3}_{-0.3}$ &  1.7$^{+0.9}_{-0.7}$ & 2.3$^{+1.6}_{-1.2}$ \\
HD~149757   & O9.2IVnn      &  31      & 3.6         & 400  & --      & -- & -- & --   \\
HD~151804   & O8Iaf         &  30    &  3.0          &  70  &   35    &  -- & -- & --  \\
HD~152247   & O9.2III       &  32     & 3.6          & 41   & 52      & 1.7$^{+0.6}_{-0.5}$ & 1.0$^{+0.4}_{-0.6}$ & 4.6$^{+3.4}_{-1.9}$  \\
HD~152249   & OC9Iab        &  31    &  3.25         &  43  &   48    &  2.8$^{+0.4}_{-0.4}$  &  1.3$^{+0.7}_{-0.7}$ & $>$7.0 \\       
HD~153426   & O8.5III       &  34    & 3.8           &  46  &    51   &  2.9$^{+0.7}_{-0.7}$ & $<1.0$ & 4.6$^{+2.6}_{-1.9}$ \\
HD~153919   & O6Iafcp       & 36     & 3.3           & 70   & 62      &   --    &   --     &  --    \\
HD~154368   & O9Iab         & 31         & 3.25      & 49   &  36     &  2.0$^{+0.7}_{-0.7}$ & 7.5$^{+3.5}_{-3.5}$ & 6.1$^{+1.4}_{-1.4}$ \\
HD~154643   & O9.7III       & 31         & 3.6       & 72   &  52     &  1.5$^{+0.8}_{-0.8}$ & $<$0.6 & 3.8$^{+1.6}_{-1.6}$  \\
HD~155806   & O7.5V((f))z   & 36         & 4.0       & 37   &  46     &  1.3$^{+0.5}_{-0.4}$ & 0.8$^{+0.2}_{-0.2}$ & 4.0$^{+2.1}_{-1.2}$ \\
HD~155889   & O9.5IV        & 33.5       & 4.0       & 33   &  25     & 0.9$^{+0.5}_{-0.4}$ & 1.3$^{+0.8}_{-0.4}$ & 3.3$^{+1.5}_{-1.2}$  \\
HD~156154   & O7.5Ib(f)     & 33.5       & 3.5       & 42   &   52    &  1.1$^{+0.2}_{-0.2}$  & 8.6$^{+3.9}_{-2.4}$ & 5.2$^{+5.8}_{-2.4}$  \\
HD~162978   & O8II((f))     & 34         & 3.5       &  35  &   46    &  1.3$^{+0.6}_{-0.3}$ & 3.9$^{+1.4}_{-0.8}$ & 3.9$^{+1.1}_{-0.7}$  \\
HD~164492A  & O7.5Vz        & 38         &  4.2      &  48  &   21    &  3.8$^{+1.3}_{-0.9}$ & 1.0$^{+0.2}_{-0.2}$ & 3.6$^{+2.5}_{-1.0}$  \\
HD~167263   & O9.5III      & 31          & 3.5       & 46        & 77      &  1.0$^{+0.6}_{-0.4}$ & $<$1.0 & 2.4$^{+1.6}_{-1.6}$            \\
HD~167264   & O9.7Iab      & 28          & 3.10      & 70        &   22    &  1.2$^{+0.4}_{-0.4}$ & 1.7$^{+0.6}_{-0.6}$ & 3.8$^{+1.7}_{-1.3}$  \\
HD~167771   & O7III(f)/O8III & 35        & 3.6       &  65       &   54    &  3.0$^{+0.7}_{-0.7}$ & 1.6$^{+0.7}_{-0.5}$ & 4.0$^{+3.1}_{-2.0}$  \\
HD~186980   & O7.5III((f))  & 35         &  3.6      &  40       &   40    &  2.5$^{+0.6}_{-0.6}$ & 3.5$^{+1.5}_{-1.2}$ & 3.4$^{+2.5}_{-1.4}$  \\
HD~188001   & O7.5Iabf      & 33         & 3.35      & 60        &   46    &  0.9$^{+0.6}_{-0.6}$ & 3.5$^{+1.8}_{-1.8}$ & 5.2$^{+3.2}_{-2.2}$  \\
HD~188209   & O9.5Iab        & 29.8      &  3.2      &  45       &   33    &  0.7$^{+0.5}_{-0.4}$ & 4.4$^{+3.6}_{-2.6}$ & 4.8$^{+3.7}_{-2.5}$  \\
HD~189957   & O9.7III       & 31         &  3.6      &  65       &   25    &  0.8$^{+0.5}_{-0.3}$ &  $<$0.7          & 2.9$^{+1.4}_{-1.0}$   \\
HD~192281   & O4.5Vn(f)     & 39         & 3.65      & 245       & --      &  1.3$^{+1.1}_{-1.0}$ & 8.4$^{+7.9}_{-4.2}$ & 1.4$^{+0.3}_{-0.3}$  \\
HD~192639   & O7.5Iabf      & 33.5       &  3.3      &  80       &  50     & 2.0$^{+1.0}_{-1.0}$  & $>$5.0 & 3.3$^{+4.9}_{-1.8}$             \\
HD~193443   & O9III         & 32         &  3.6      &  46       &   60    & 2.5$^{+1.2}_{-1.0}$  &  $<$0.7  & 4.57*                      \\
HD~199579   & O6.5V((f))z   & 41.5       &  4.15     &  55       &   50    &  2.4$^{+1.4}_{-0.9}$ & 0.8$^{+0.3}_{-0.2}$ & 6.0$^{+2.8}_{-2.5}$   \\
HD~201345   & ON9.2IV       & 34         &  4.0      &  75       &   32    &  $<$0.4     &  4.0$^{+2.2}_{-1.3}$ & 3.6$^{+2.7}_{-2.2}$        \\
HD~203064   & O7.5IIIn((f)) & 34         & 3.6       & 300       & --      & 1.6$^{+0.4}_{-0.4}$ & 1.6$^{+0.7}_{-0.7}$ & 4.57*               \\
HD~206183   & O9.5IV-V      & 33         &  4.1      &  15       &  6      & 1.2$^{+0.3}_{-0.2}$ & 0.7$^{+0.3}_{-0.3}$ & 2.8$^{+1.3}_{-1.1}$    \\
HD~207198   & O8.5II        & 32.5       &  3.50     &  60       &  27     & 1.4$^{+0.3}_{-0.3}$ & 1.9$^{+0.4}_{-0.3}$ & 2.8$^{+1.2}_{-0.7}$   \\  
HD~207538   & O9.7IV        & 30.5       &  3.75     &  20       &   27    & 0.8$^{+0.7}_{-0.4}$ & 1.1$^{+0.7}_{-0.4}$ & 2.8$^{+1.1}_{-0.9}$   \\
HD~209975   & O9Ib          & 30.5       &  3.35     &  48       &   40    & 1.2$^{+0.4}_{-0.4}$ & 2.1$^{+1.0}_{-0.7}$ & 5.5$^{+3.4}_{-2.5}$   \\
HD~210809   & O9Iab         & 30.5       &  3.35     &  57       &   50    & 0.7$^{+0.4}_{-0.3}$ & 3.6$^{+2.0}_{-0.9}$ & 3.4$^{+2.8}_{-1.6}$   \\
HD~210839\tablefootmark{a}& O6I(n)fp & 36  & 3.5     &  210      &  80     & 0.6$^{+0.2}_{-0.2}$ & 6.0$^{+1.9}_{-1.3}$ & 2.5$^{+2.0}_{-2.0}$  \\
HD~214680   & O9V           & 35.0       &  4.05     &  15       &  15     & 2.7$^{+1.3}_{-1.1}$ & 1.6$^{+0.8}_{-0.6}$ & 6.0$^{+3.4}_{-2.5}$   \\ 
HD~218195   & O8.5III Nstr  & 34         &  3.8      &  34       &  34     & 2.0$^{+0.4}_{-0.4}$ & 5.0$^{+4.0}_{-2.0}$ & 4.6$^{+4.1}_{-2.9}$   \\
HD~218915   & O9.2Iab       & 30         &  3.25     &  50       &  32     & 0.9$^{+0.5}_{-0.5}$ & 5.6$^{+1.8}_{-1.8}$ & 4.0$^{+3.2}_{-2.3}$   \\
\hline
\end{tabular}
\tablefoot{(a) Adopted from \citet{jc12}. Spectral types are from \citet{sota11,sota14}.}
\end{center}
\end{table*}

\setcounter{table}{0}

\begin{table*}
\begin{center}
\caption{Continued.} \label{tab_param2}
\begin{tabular}{lccccccccc}
\hline
Star        & ST            &   Teff     &   logg    &  \vsini\  &  \vmac  &  C/H         &    N/H      &    O/H      \\    
            &               & [kK]       &           &  [\kms]   &  [\kms] &  [10$^{-4}$]  &  [10$^{-4}$] &  [10$^{-4}$]  \\
\hline
\smallskip
HD~227757   & O9.5V         & 34         & 4.0       & 21        & 13      & 2.6$^{+0.6}_{-0.5}$ & 0.8$^{+0.5}_{-0.3}$ & 5.6$^{+2.9}_{-2.1}$    \\
HD~258691   & O9V           & 33.5       & 4.0       & 14        & 18      & 1.2$^{+0.5}_{-0.5}$ & $<$1.0 & 3.0$^{+1.3}_{-1.3}$ \\
HD~328856   & O9.7II          & 30       & 3.4       &  58       &   55    &  1.8$^{+0.6}_{-0.3}$ & 1.5$^{+0.8}_{-0.7}$ & 5.1$^{+2.9}_{-2.1}$    \\ 
BD-134930   & O9.5V           & 33       & 4.1       &  $<$15    &   $<$5  &  1.5$^{+0.9}_{-0.7}$ & 0.9$^{+0.7}_{-0.4}$ & 4.3$^{+2.2}_{-1.8}$   \\
BD+60499    & O9.5V           & 34       & 4.0       &  30       &   25    &  2.5$^{+1.0}_{-1.0}$ & 0.8$^{+0.3}_{-0.3}$ & 4.4$^{+2.3}_{-1.6}$    \\
 & & & & & \\
\hline
Magnetic stars \\
\hline
HD~108     &  O8f?p         & 35       & 3.5       & 1    &  38    & 1.5$^{+0.5}_{-0.5}$ &  3.0$^{+1.1}_{-0.8}$ & 2.3$^{+2.0}_{-0.9}$  \\
HD~57682   &  O9.2IV        & 34.5     & 4.0       & 23   &  14    & 1.0$^{+0.5}_{-0.3}$ &  1.2$^{+0.8}_{-0.5}$ & 3.8$^{+2.3}_{-2.1}$  \\
HD~148937  &  O6f?p         & 40       & 4.0       & 22   &  60    & 1.5$^{+0.5}_{-0.4}$ &  5.0$^{+1.2}_{-1.0}$ & 1.7$^{+3.0}_{-0.9}$  \\
HD~191612  &  O8f?p         & 36       & 3.75      & 1    &  38    & 0.7$^{+0.3}_{-0.3}$ &  2.0$^{+0.8}_{-0.8}$ & 1.1$^{+1.8}_{-0.6}$  \\
$\theta^1$Ori~C & O7V       & 38       & 4.2       & 24   &  32    & 1.4$^{+0.6}_{-0.4}$ & 0.45$^{+0.25}_{-0.2}$ & 1.0$^{+0.5}_{-0.5}$  \\
CPD-28 2561 & O6.5f?p       & 35       & 4.0       & 1    &  70    & 0.5$^{+0.3}_{-0.3}$ & $<$0.5             & $<$1.5            \\ 
Tr16 22    &  Of?p          & 35       & 4.2       & 38   &  9     & 2.2$^{+1.0}_{-1.0}$ & $<$0.8             & 4.57*            \\
\hline
\end{tabular}
\tablefoot{Spectral type of BD-134930 from Hillenbrand et al. (1993). Symbol * stands for 'adopted'.}
\end{center}
\end{table*}

%%%%%%%%%%%%%%%%%%%%%%%%%%%%%%%%%%%%%%%%%%%%%%%%%%%%%%%%%%%%%%%%%%%%%%%%%%%%%%%%%%%%%%%%%%%%%%%%%%%%%%%%%%%%%%%%%%%%%%%%%%%%%%%
%%%%%%%%%%%%%%%%%%%%%%%%%%%%%%%%%%%%%%%%%%%%%%%%%%%%%%%%%%%%%%%%%%%%%%%%%%%%%%%%%%%%%%%%%%%%%%%%%%%%%%%%%%%%%%%%%%%%%%%%%%%%%%%
\section{Surface abundances}
\label{s_ab}

%---------------------------------------------------------------
\subsection{General trends}
\label{s_ab_obs}
      
\citet{maeder09} stressed that surface abundances depend on several parameters in single stars: rotation, mass, metallicity. We are studying Galactic stars. To first order, we will assume that they all have a solar metallicity. This may not be true for a few objects, but since the sample is biased towards relatively bright stars (for magnetic studies), it is reasonable to assume that most of the targets are close to the Sun and thus share its metallicity. As already stressed, our sample contains mainly stars with initial masses between 20 and 50 \msun. Although wide, this range is not extreme. We will investigate the surface abundances of sub-samples in the following, but we first present the results for the entire sample. 

\begin{figure}[t]
\centering
\includegraphics[width=9cm]{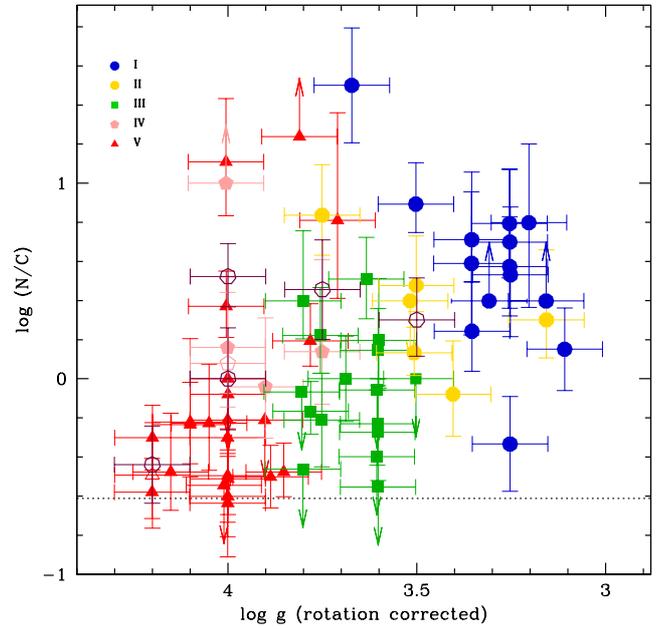}
\caption{log N/C (by number) as a function of \logg\ for the sample stars. The dotted line shows the solar value of log(N/C) according to \citet{ga10}.}
\label{fig_CN}
\end{figure}

Fig.\ \ref{fig_CN} displays the values of log (N/C) as a function of surface gravity. N/C traces the evolution in terms of nucleosynthesis, while \logg\ indicates to first order the evolution in the HR diagram and thus the age. In this figure, the dwarfs, giants and supergiants are very well separated in surface gravities. The main feature of the plot is the clear and significant increase of N/C as \logg\ decreases. All (but five) dwarfs have log(N/C)$<$0 and most of them have log(N/C) consistent with -0.6, the solar value \citep{ga10}. Hence, as expected from their luminosity class, most dwarfs are barely evolved both in terms of surface gravity and surface abundances. The dwarfs with log(N/C)=0.4--1.1 are HD~192281, HD~46223 and HD~46150, the earliest dwarfs of our sample (spectral types O4-O5), and thus also the most massive ones. Since chemical enrichment is expected to be stronger in more massive stars, finding them above the bulk of O dwarfs in Fig.\ \ref{fig_CN} is natural. We will come back to this below. The dwarf with log(N/C)=0.2 is HD~36879. Its spectral type is O7. It is thus intermediate between the early O dwarfs and the bulk of the O dwarfs, made of O8-O9.5 stars.  
Only HD~14633, with log(N/C)=1.24, escapes the expected trends. It is a late O dwarf (O8.5) but shows an extreme enrichment. Interestingly, it is classified as an ON star, indicating that its nitrogen lines are especially strong. This is confirmed by our analysis: HD~14633 is particularly nitrogen rich. 

The giant stars in Fig.\ \ref{fig_CN} show a wider range of N/C. If the lowest values are also consistent with the solar value, N/C can reach 3 (log (N/C) = 0.5). Quantitatively, giants are thus on average more chemically evolved than dwarfs.

Finally, supergiants confirm the trend seen with giants: the lower the surface gravity, the larger the N/C ratio. For supergiants, we find log(N/C) to be systematically above 0, reaching values of about 1.0. HD~66811, with log (N/C) = 1.5, stands above the bulk of supergiants because its initial mass is higher (see below). HD~152249, a OC9Iab star, is the only clear outlier with a negative value of log(N/C). Qualitatively, we thus observe a clear trend of chemical enrichment as surface gravity decreases, and thus as evolution proceeds. 
Although less numerous, subgiants and bright giants do not depart from this trend\footnote{The subgiant with log (N/C) = 1.0 is also an ON star.}. 
We can conclude from Fig.\ \ref{fig_CN} that in the 20-50 \msun\ mass range the ratio of nitrogen over carbon surface abundances increases as surface gravity decreases.

\begin{figure}[t]
\centering
\includegraphics[width=9cm]{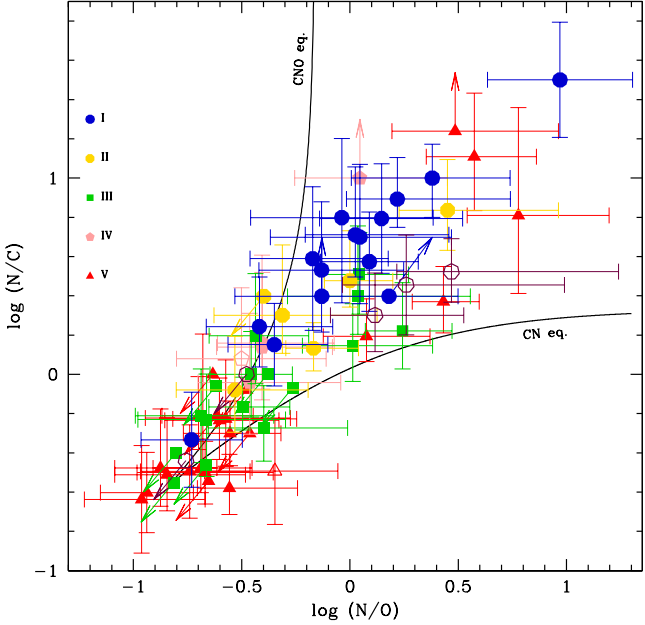}
\caption{log (N/C) (by number) as a function of log (N/O) for the sample stars. The expected trends for the case of CN and CNO equilibrium are shown by the solid lines.}
\label{fig_CNO}
\end{figure}

Fig.\ \ref{fig_CNO} provides further insight into the chemical evolution of the sample stars. The ratio of nitrogen to carbon abundances is shown as a function of the ratio of nitrogen to oxygen abundances (in a log-log diagram so that the full range of values can be seen at once). There is a remarkable correlation between both ratios: the larger N/O, the larger N/C. This is in excellent agreement with the expectations of nucleosynthesis through the CNO cycle.
\citet{maeder14} studied quantitatively the relation between the N/C and N/O ratio in two cases. First, for the most massive stars where the CN cycle reaches equilibrium immediately and the $^{12}$C abundance is constant, they find that 

\begin{equation}
\frac{d(N/C)}{d(N/O)}=\frac{N/C}{N/O}\frac{1}{1+N/O}
\end{equation}

\noindent which we can rewrite as

\begin{equation}
\frac{dlog(N/C)}{dlog(N/O)}=\frac{1}{1+N/O}
\label{eq_cno1}
\end{equation}

\noindent In the second case, which applies to stars with masses of the order 5-15 \msun, it is the $^{16}$O abundance which remains constant and the following relation is obtained

\begin{equation}
\frac{d(N/C)}{d(N/O)}=\frac{N/C}{N/O}(1+N/C)
\end{equation}
 
\noindent which leads to 

\begin{equation}
\frac{dlog(N/C)}{dlog(N/O)}=1+N/C
\label{eq_cno2}
\end{equation}

\noindent We have included these limiting cases in Fig.\ \ref{fig_CNO} by means of black solid lines.  Remarkably, all stars lie in between these lines (within the error bars). The possible small offset towards lower values of log(N/O) can easily be accounted for by a slightly different solar N/O ratio, within the uncertainties of the solar abundances measurements. Finding stars in between both lines clearly shows that they are in a state where neither CN nor CNO equilibrium is dominant. The dispersion of N/C for a given N/O is a natural consequence of various internal conditions favouring either the CN or the CNO cycle.

Fig.\ \ref{fig_CNO} provides additional information regarding chemical evolution. Looking again at the three groups defined by dwarfs, giants and supergiants, one sees a very clear separation between each group. Dwarfs are the least evolved (except for a few objects), then come the giants with intermediate values of N/C and N/O and finally supergiants with the largest ratios. This is the first time such a clear evolutionary sequence between different luminosity classes is observed for O stars.
The evolution of chemical abundances can be directly related to the evolution in terms of spectral appearance and surface gravity. This conclusion applies in the mass range probed by our analysis (20-50 \msun) and is valid for an ensemble of stars. There are a few outliers deviating from this general conclusion. In the dwarfs sub-sample, the earliest O stars (HD~46150, HD~46223 and HD~192281) still stand out, as in Fig.\ \ref{fig_CN}. The higher mass of these objects is the natural explanation (see below). HD~36879 -- O7V, at log(N/O)=0.1 -- is more evolved than the late-type O dwarfs, but not as evolved as the earliest objects. This is a qualitative confirmation of the expectation that more massive stars are more mixed than lower mass stars \citep{maeder09}. HD~14633 is the only clear outlier among dwarfs: it is the second most enriched object (and the carbon abundance being only an upper limit, it may be an extraordinary object). In a similar way, the other ON star of our sample -- the subgiant HD~201345 -- also stands out: its enrichment is equivalent to that of the most evolved supergiants. Clearly, the two ON stars bear peculiar properties. We postpone a detailed analysis of this peculiar class of objects to a subsequent publication. Finally, the last outlier is HD~152249, a supergiant barely chemically evolved and member of the OC class. The most chemically enriched object of our sample is HD~66811, the most massive O supergiant. Given its initial mass and its advanced evolutionary status (in terms of temperature and gravity) it is not surprising that it is also the most chemically evolved. HD~66811 is also suspected to be a runaway star which may indicate past binary interaction.

\begin{figure}[t]
\centering
\includegraphics[width=9cm]{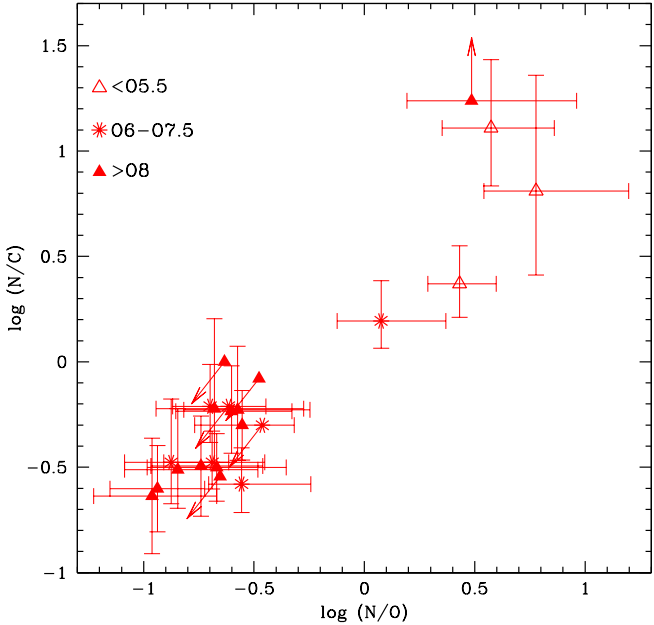}
\caption{log (N/C) (by number) as a function of log (N/O) for the dwarfs. Different symbols refer to different spectral type bins, as indicated in the figure. Symbols are the same as in Fig.\ \ref{fig_CN}.}
\label{fig_CNO_V}
\end{figure}

Luminosity class V is the class containing the largest number of object of our sample. They also cover a wider range of spectral type (O4 to O9.7) and thus a wide mass range. In Fig.\ \ref{fig_CNO_V} we show the N/C and N/O ratios for dwarfs. The objects have been separated in three spectral type bins: earlier than O5.5, later than O8, and intermediate. The early O dwarfs are significantly more mixed than the late O stars. Most intermediate O dwarfs have CNO abundances similar to late-type stars. The only exception (HD~36879) shows an intermediate degree of mixing. Among a given spectral class, early-type stars are more massive than late stars. Anticipating on the discussion of the next section, Fig.\ \ref{fig_cn_mod} shows that early O dwarfs, which have \teff\ hotter than 39000~K (see Table \ref{tab_param2}) have masses in the range 35-60 \msun\ depending on the evolutionary tracks. Late O stars, \teff\ $<$ 34000 K, have masses between 18 and 25 \msun. Hence, Fig.\ \ref{fig_CNO_V} clearly shows that mixing is stronger in more massive stars. This is one of the predictions of evolutionary models including rotation \citep[e.g. Fig. 1 of][]{maeder09}. 

\vspace{0.4cm}

Our study of surface chemical abundances thus demonstrates that single Galactic O stars show the products of nucleosynthesis occurring through the CNO cycle at their surface. In addition, mixing is stronger when 1) stars evolve off the main sequence and 2) at higher masses. These results are fully consistent with stellar evolution of rotating stars. However, a crucial test of such models would be to show that faster rotators are more enriched than slower ones. This would have to be done in narrow mass and luminosity class ranges to isolate the effects of rotation from those described above. Unfortunately, our sample is still too small to allow such a comparison. Although it includes stars with \vsini\ $<$ 30 \kms and others with \vsini\ $>$ 150 \kms, the number of fast rotators is small (9 objects) and spread over the entire mass range (20 to 50 \msun). In a given mass bin, the number of fast rotators is thus too small to lead to any conclusive results. This is a clear direction for future work: analyze a larger sample of fast rotators.

%---------------------------------------------------------------
\subsection{Model predictions}
\label{s_ab_mod}

\begin{figure*}%[tpb]
     \centering
     \subfigure{
          \includegraphics[width=.42\textwidth]{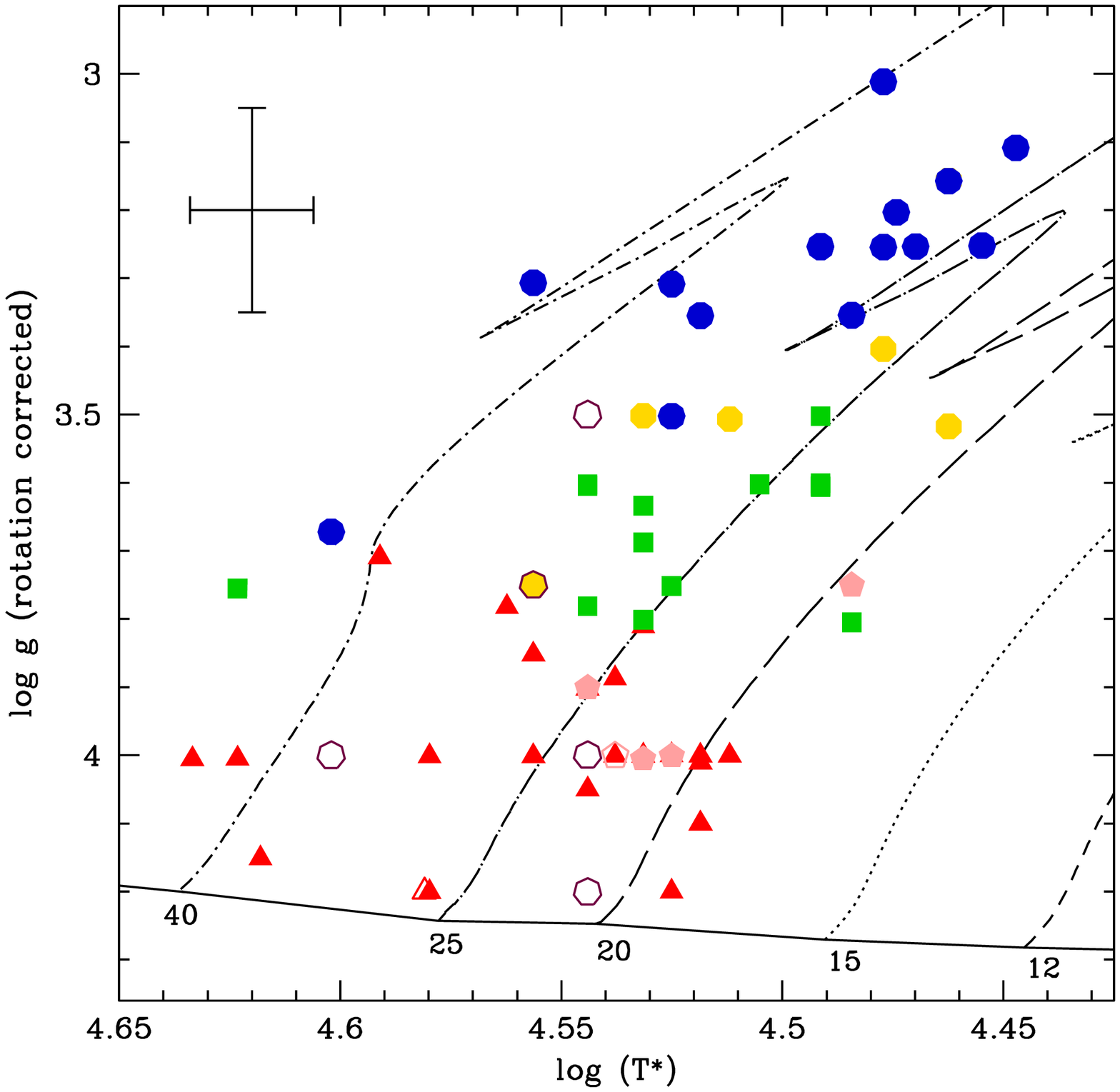}}
     \hspace{0.2cm}
     \subfigure{
          \includegraphics[width=.42\textwidth]{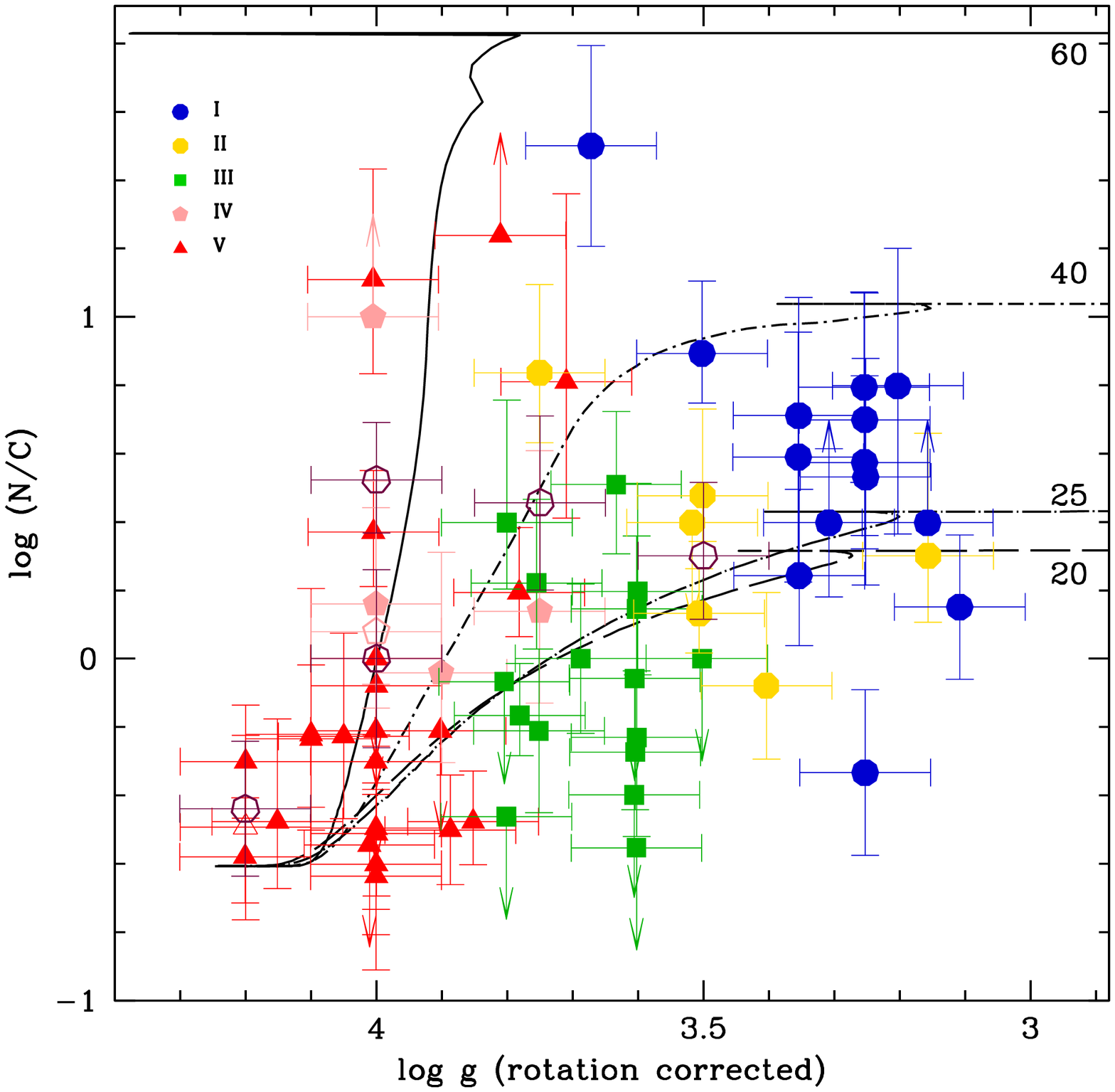}}\\
     \subfigure{
          \includegraphics[width=.42\textwidth]{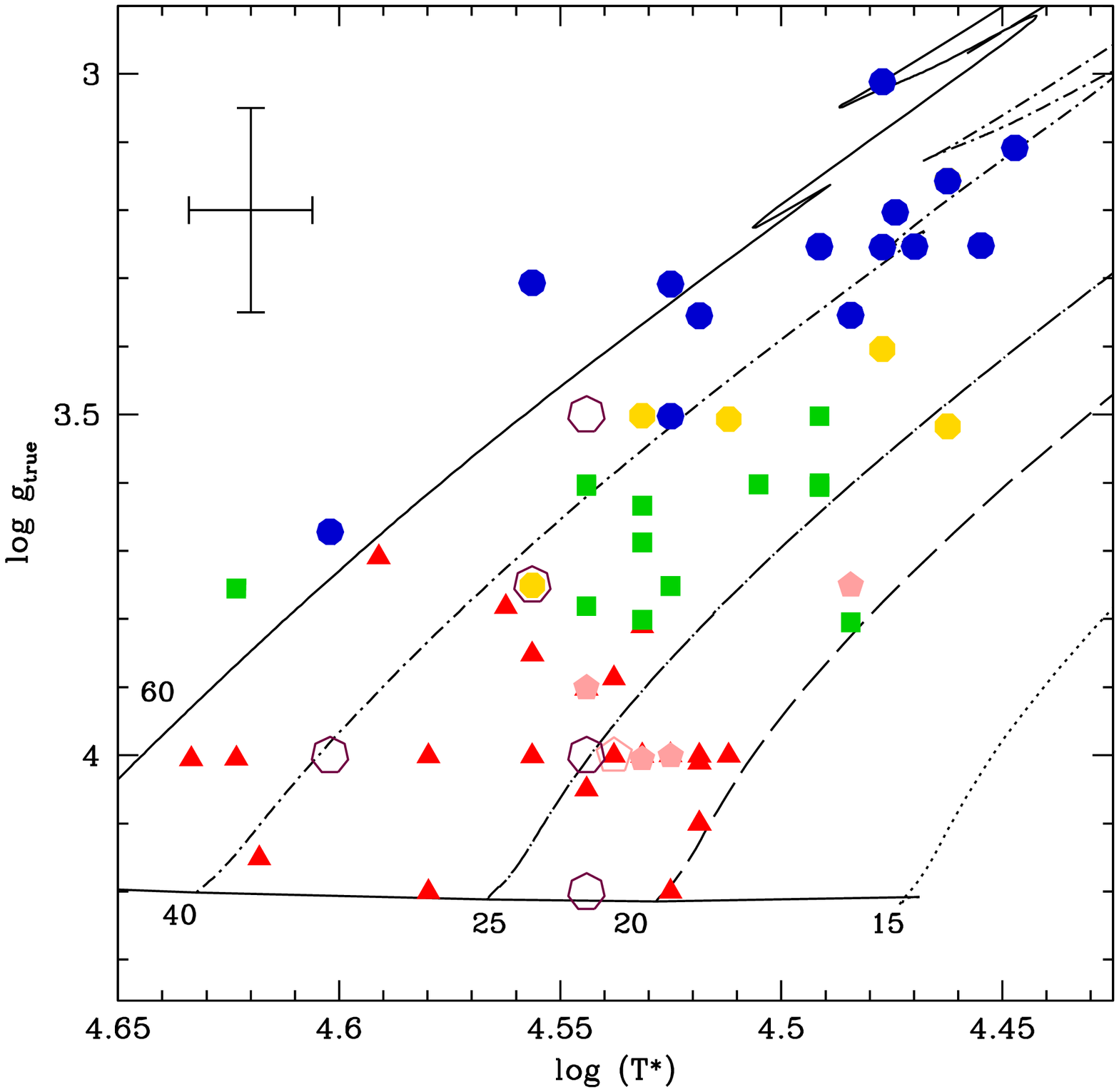}}
     \hspace{0.2cm}
     \subfigure{
          \includegraphics[width=.42\textwidth]{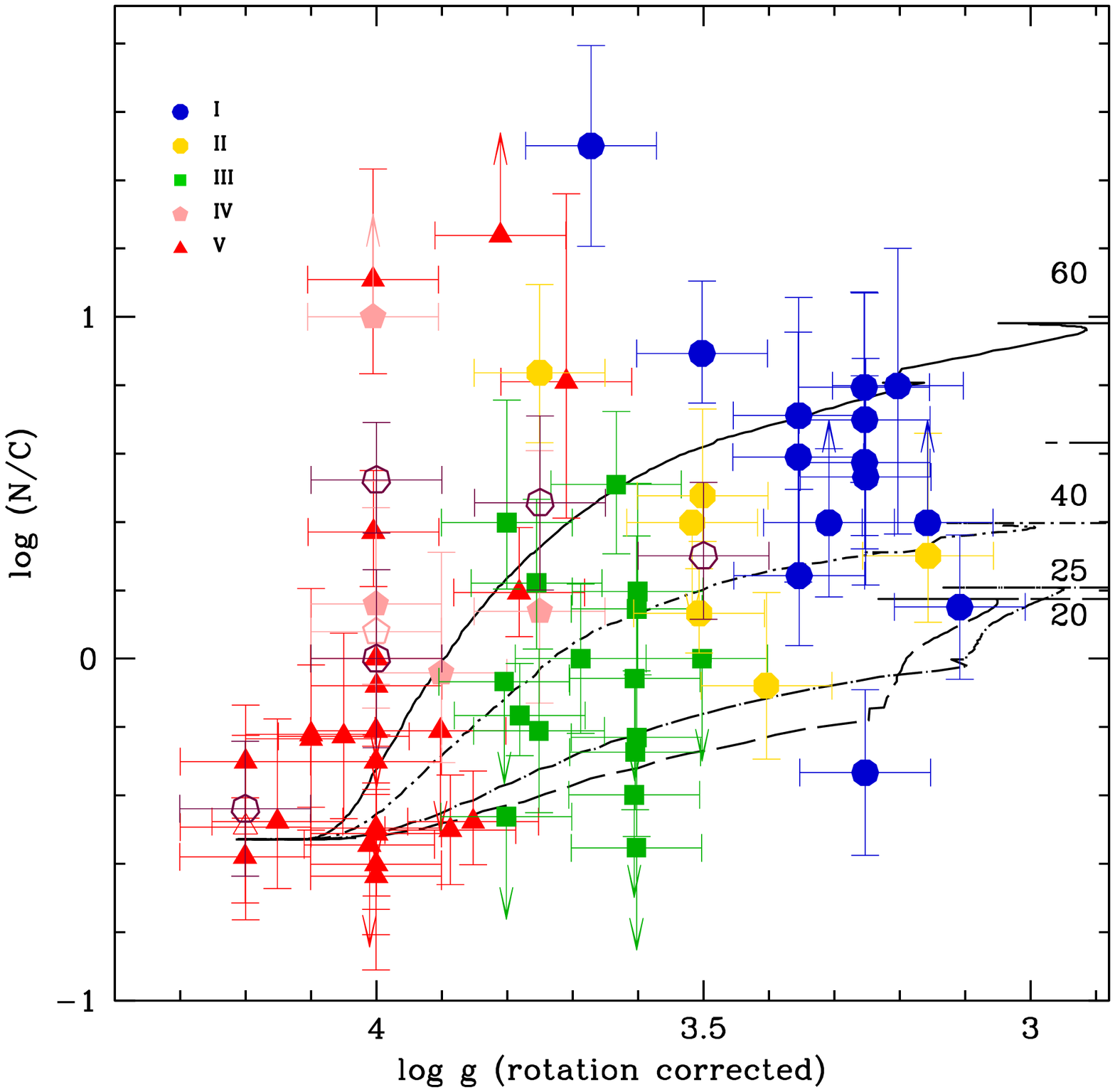}}\\
     \subfigure{
          \includegraphics[width=.42\textwidth]{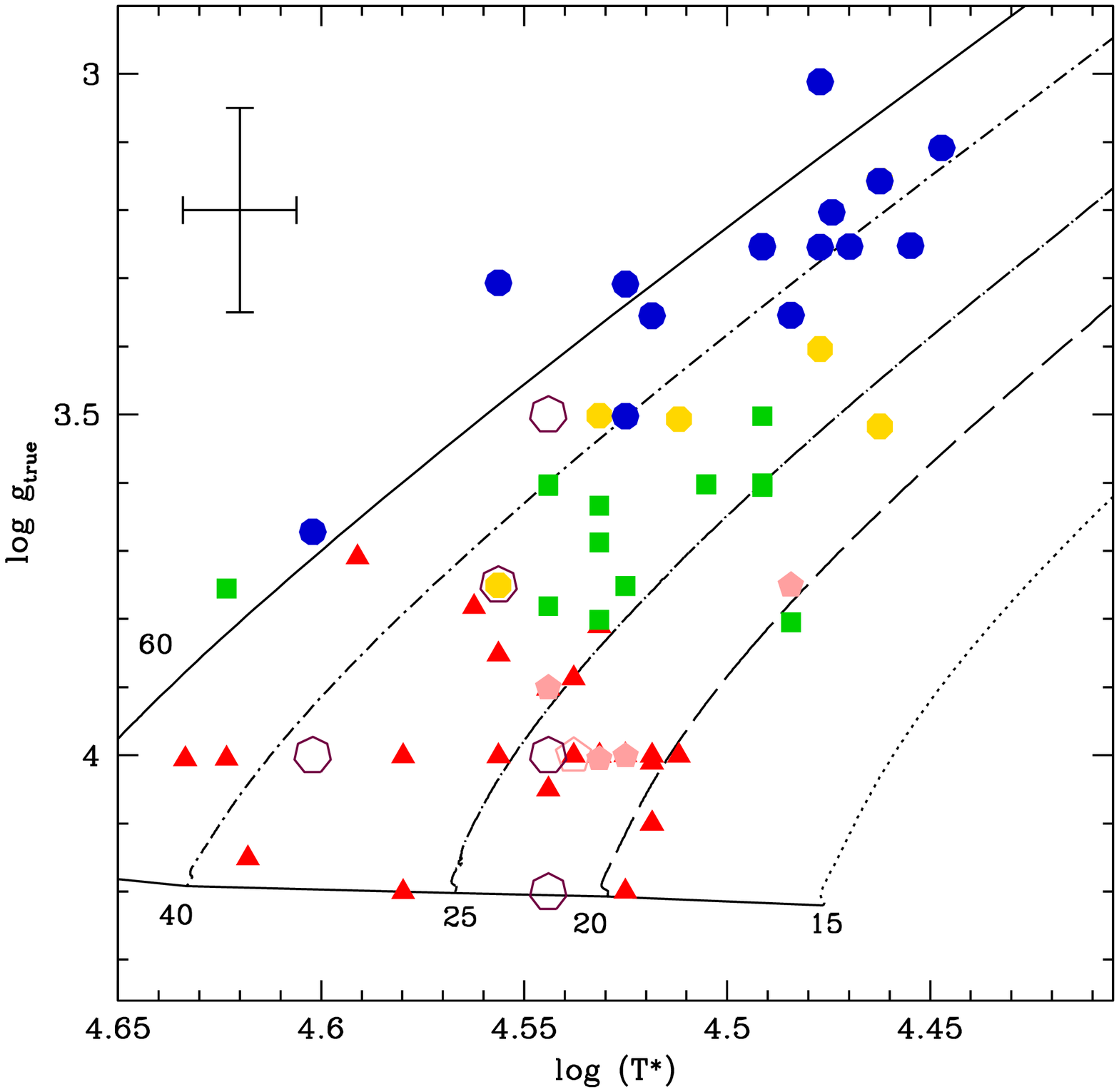}}
     \hspace{0.2cm}
     \subfigure{
          \includegraphics[width=.42\textwidth]{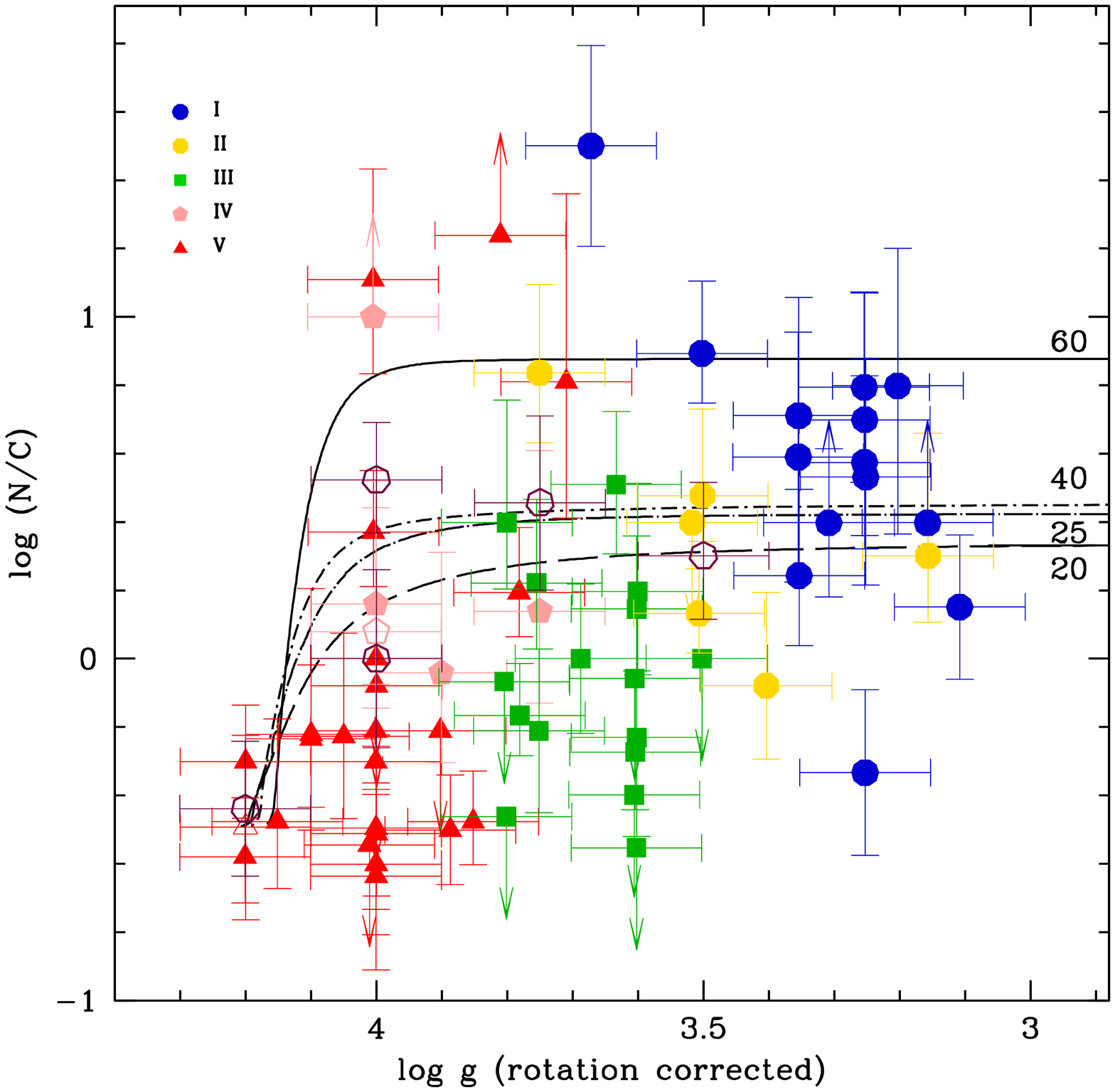}}
     \caption{\textit{Left panels}: \logg\ -- \teff\ diagram with the sample stars. \textit{Right panels}: log(N/C) -- \logg\ diagram. In each panel, evolutionary tracks for different initial masses (indicated in the figures) are overplotted. The upper, middle, lower panels corresponds to tracks from \citet{ek12}, \citet{cl13}, \citet{brott11} respectively.}
     \label{fig_cn_mod}
\end{figure*}

Several grids of models for massive stars have been published in the last years. In this section, we compare our results to the predictions of \citet{brott11}, \citet{ek12} and \citet{cl13}. All three grids include rotation (with different prescriptions) with an initial velocity of about 300 \kms. \citet{brott11} also include magnetic field. Metallicity is set to solar (Z=0.014 for Ekstr$\ddot{o}$m et al., Z=0.01345 for Chieffi \& Limongi) except for the grid of \citet{brott11} for which Z=0.0088 (see \cite{mp13} for a discussion on this topic). 

Fig.\ \ref{fig_cn_mod} shows two diagrams for each grid of models: on the left part, the \logg\ -- \teff\ diagram; on the right part, the log(N/C) -- \logg\ diagram. In each figure, evolutionary tracks for various initial masses are overplotted. Let us begin with the upper panels corresponding to the models of \citet{ek12}. The \logg\ -- \teff\ diagram reveals that most stars have initial masses between 20 and 40 \msun. In the log(N/C) -- \logg\ diagram, as we have already described, most dwarfs lie in the lower left corner, at high \logg\ and low log(N/C). Given the error bars, all tracks (with any initial mass) are consistent with the position of these objects which are essentially chemically unevolved. Among the dwarfs showing chemical mixing at their surface, only HD~46150 (\logg\ = 4.0 and log(N/C) = 0.37) and HD~46223 (\logg\ = 4.0 and log(N/C) = 1.10) are more evolved than expected: their enrichment is similar to that of a 60 \msun\ star while their initial mass is closer to 40-50 \msun. The supergiant HD~66811 has an initial mass above 40 \msun\ and it is the most evolved of all our targets. In the log(N/C) -- \logg\ diagram it is located between the 40 and 60 \msun\ track and is thus well reproduced by the Geneva tracks. The other supergiants for which we have determined surface abundances are all located in between the 25 \msun\ and the 40 \msun\ tracks in the left panel. Their N/C ratios are consistently reproduced by these tracks in the right panel, with only one outlier and one star marginally explained. The outlier (HD~152249 - OC star) is much less evolved than expected for its mass and surface gravity. The other possible outlier is HD~167264 with log(N/C)=0.15. The enrichment is slightly weaker than expected, but the difference is marginal and could be easily explained by a slower rotation. Finally giants, which cluster around the 25 \msun\ track in the left panel (except HD~93250, above 40 \msun), are on average also correctly reproduced by this track in the log(N/C) -- \logg\ diagram. However, six objects appear to be less evolved than expected. Given the overall success of the Geneva tracks to explain the properties of the entire sample, we attribute this to rotation effects. The two ON stars are, together with the OC star HD~152249, HD~46150 and HD~46223, the only true outliers in the right panel: their strong enrichment cannot be explained by the tracks near which they lie in the left panel. They are obviously more evolved. The main conclusion is that the Geneva tracks can consistently reproduce the effective temperature, surface gravity and surface N/C ratio of 90\% (80\% excluding the weakly enriched giants) of the sample. 

Let us now turn to the middle panels of Fig.\ \ref{fig_cn_mod}. The evolutionary tracks are from \citet{cl13}. We see in the left panel that most stars have initial masses between 20 and 60 \msun. The range is wider than in the case of the Geneva models because the Chieffi \& Limongi tracks evolve more rapidly towards lower effective temperature. Consequently, if the conclusions regarding the initial masses of the dwarfs is unchanged, almost all supergiants are now in the 40-60 \msun range, and giants have masses between 25 and 40 \msun. As for the Geneva models, in the log(N/C) -- \logg\ diagram, the dwarfs are mostly accounted for by the 20-40 \msun\ tracks. HD~46150 and HD~46223 are more enriched than predicted by the 60 \msun\ track below which they lie in the right panel. HD~66811 is also much too enriched (as are the two ON stars). The supergiants are reasonably well accounted for by the 40 and 60 \msun\ track between which they lie in the \logg\ -- \teff\ diagram. HD~152249 is less enriched than expected. HD~156154, with \logg\ = 3.5 and log(N/C)=0.89 is more enriched than expected from the 40 \msun\ track (on which it perfectly falls in the left panel). The bright giant HD~34656 (\logg\ = 3.75 and log(N/C)=0.83) faces similar problems with the same track. On average, the chemical enrichment of the giants is very well reproduced by the 25 and 40 \msun\ tracks. The only clear exceptions are HD~218195 (\logg\ = 3.5 and log(N/C)=0.4) and HD~24912 (\logg\ = 3.6 and log(N/C)=0.5): their enrichment is higher than expected for 25 \msun\ stars. Two giants with \logg\ = 3.6 also have upper limits on N/C that cannot be explained by the 25 \msun\ track. Overall, the models by \citet{cl13} are able to reproduce the observed properties of 80\% of the sample. 

Finally, let us move to the lower panels of Fig.\ \ref{fig_cn_mod} where the tracks by \citet{brott11} are used. The \logg\ -- \teff\ diagram is very similar to that obtained when using the Chieffi \& Limongi tracks, hence the conclusions regarding the initial masses of the sample stars are the same. In the right panel, the evolution is much faster than in the two previous grids: the N/C ratio reaches a maximum at \logg\ $\sim$ 4.0, immediately after the zero age main sequence, and remains constant for the rest of the evolution. Consequently, the N/C values of only about half of the dwarfs can be explained. The majority of the giants are less enriched than observed. Only the three stars with N/C $>$ 1.0 are consistent with the 20-40 \msun\ tracks. For supergiants, the situation is better since the range of observed N/C is compatible with the predictions of the 40 and 60 \msun\ tracks. HD~66811, HD~46223 and the ON stars are more evolved than expected. We performed the analysis with a lower initial velocity (200 \kms). The general behaviour of the tracks in the log(N/C) -- \logg\ diagram is the same, except that the plateaus are shifted to lower values (the \logg\ -- \teff\ diagram is barely changed). This provides a better explanation for the giants, but the supergiants are no longer accounted for. The main conclusion regarding the Brott et al.\ tracks is that they reproduce only 50\% of the sample stars. The reason is the very fast mixing in the early phases of evolution.

The comparison of the properties and chemical enrichment of the sample stars with the three grids of models has shown that the Ekstr$\ddot{o}$m et al.\ and Chieffi \& Limongi tracks do a similar job in reproducing the observed trends. Their success rate is good. However, there also important differences regarding the initial mass of the objects inferred from either grid. The masses obtained from the Chieffi \& Limongi are higher than those obtained from the Geneva grid. The reason is the difference in luminosity and effective temperature, and thus radius and \logg, for the same initial mass between both grids. To discriminate between both sets of models, one would need independent mass determinations for the sample stars or alternatively constraints on the radius and thus the luminosity. Once accurate distances are known, as it will be the case in a few years with the \textit{Gaia} mission, such tests will be possible. The Brott et al.\ grid of models appears to be less suited to explain the properties of the sample stars. Mixing is too strong in this grid, which is at least partly due to the different treatment of transport of chemical elements compared to the two other grids (see Martins \& Palacios (2013) for a summary of the input physics in different codes).

%---------------------------------------------------------------
\subsection{Comparison with other studies: mass and metallicity effects}
\label{s_ab_comp}

\begin{figure}[t]
\centering
\includegraphics[width=9cm]{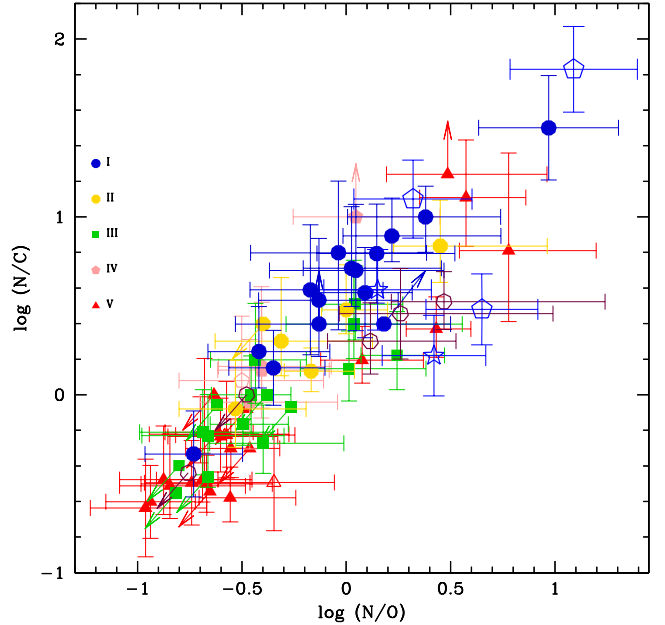}
\caption{log (N/C) as a function of log (N/O) for the stars of the present study and the O supergiants of \citet{jc12} -- open stars: spectral type O6-O7.5; open pentagons: spectral type O4-O4.5. The supergiants with log (N/C) $>$ 1.0 are all of spectral type O4.}
\label{fig_cno_jc12}
\end{figure}

The analysis of abundances in O stars is a quite recent subject. If nitrogen has been the focus of most studies so far \citep{hunter08,martins12,rg12}, abundances for carbon and oxygen are still quite rare. This is mainly due to the requirement for high quality data and state-of-the-art models to perform their analysis. In Fig.\ \ref{fig_cno_jc12} we show the log (N/C) -- log (N/O) diagram where the O supergiants studied by \citet{jc12} have been added (open symbols). The analysis of these eight objects was performed with the same tools. The abundances were derived from UV spectroscopy rather than optical spectroscopy. HD~66811 and HD~210839 are part of both our and Bouret et al.'s sample. For HD~66811 we found that the parameters derived by Bouret et al. gave an excellent fit to our optical spectrum and thus adopted their parameters for our study. For HD~210839, we derived slightly different abundances, still compatible with those of Bouret et al.\ within the error bars. The sample of \citet{jc12} contained four O4 stars, one O4.5 stars, and three O6 to O7.5 stars. The spectral types of the supergiants of our study range between O7.5 and O9.7 (if one excludes HD~66811 and HD~210839). The two sample are thus complementary, probing different mass ranges: our supergiants have initial masses below 40 \msun, those of Bouret et al.\ have masses higher than 40 \msun\ (based on the Geneva models). Fig.\ \ref{fig_cno_jc12} reveals two things: first, the O6-O7.5 stars of Bouret et al.\ have about the same N/C and N/O values as the most evolved of our supergiants; second, the O4 stars of Bouret et al.\ are more chemically enriched than the mid-to-late type supergiants. These results teach us that in the supergiant sample, more massive stars show a higher degree of chemical mixing. This latter result is in agreement with our findings for Galactic dwarfs (see Sect.\ \ref{s_ab_obs}).

\begin{figure*}[]
     \centering
\subfigure{ \includegraphics[width=.45\textwidth]{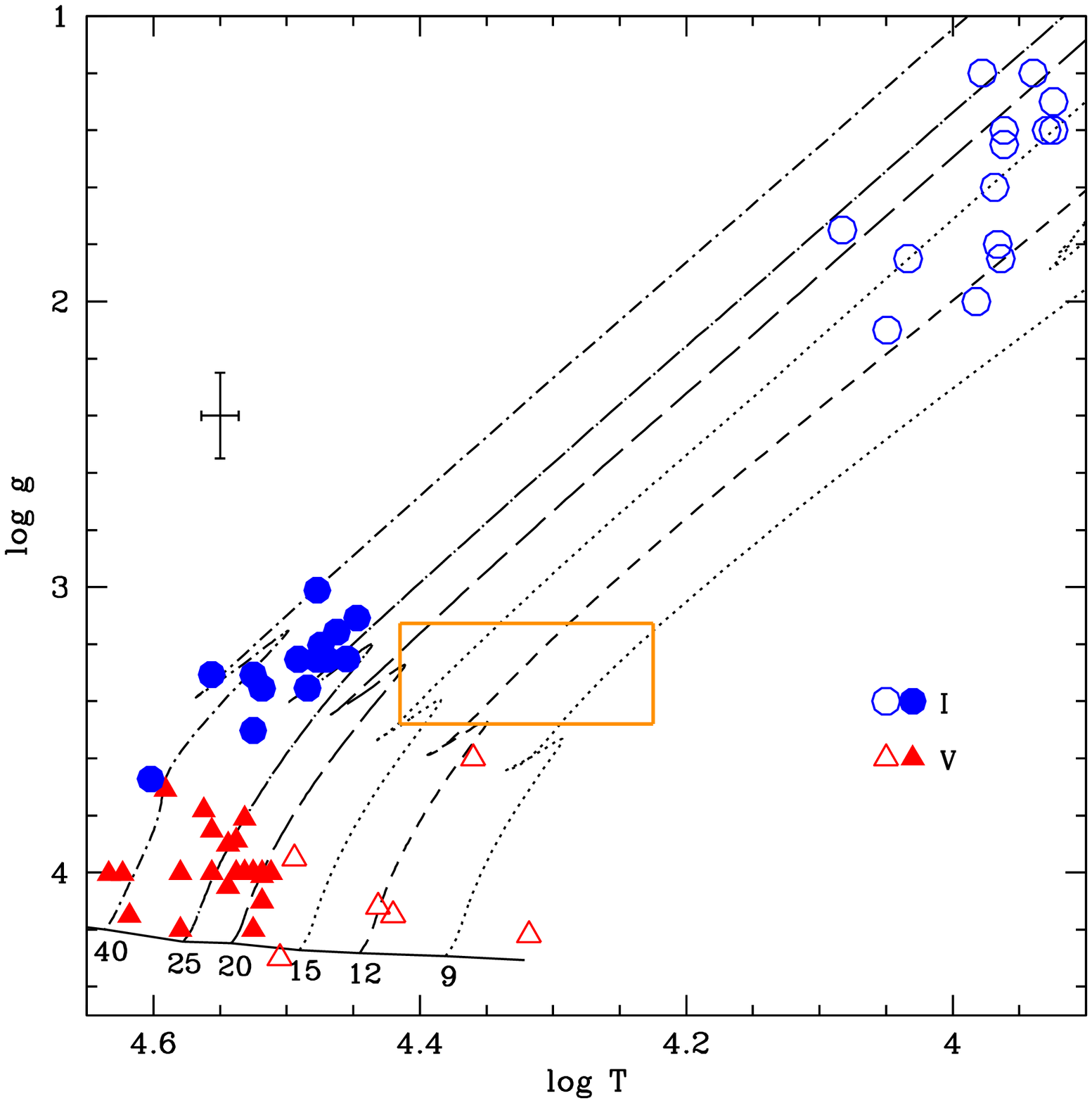}}
     \hspace{0.1cm}
     \subfigure{
           \includegraphics[width=.45\textwidth]{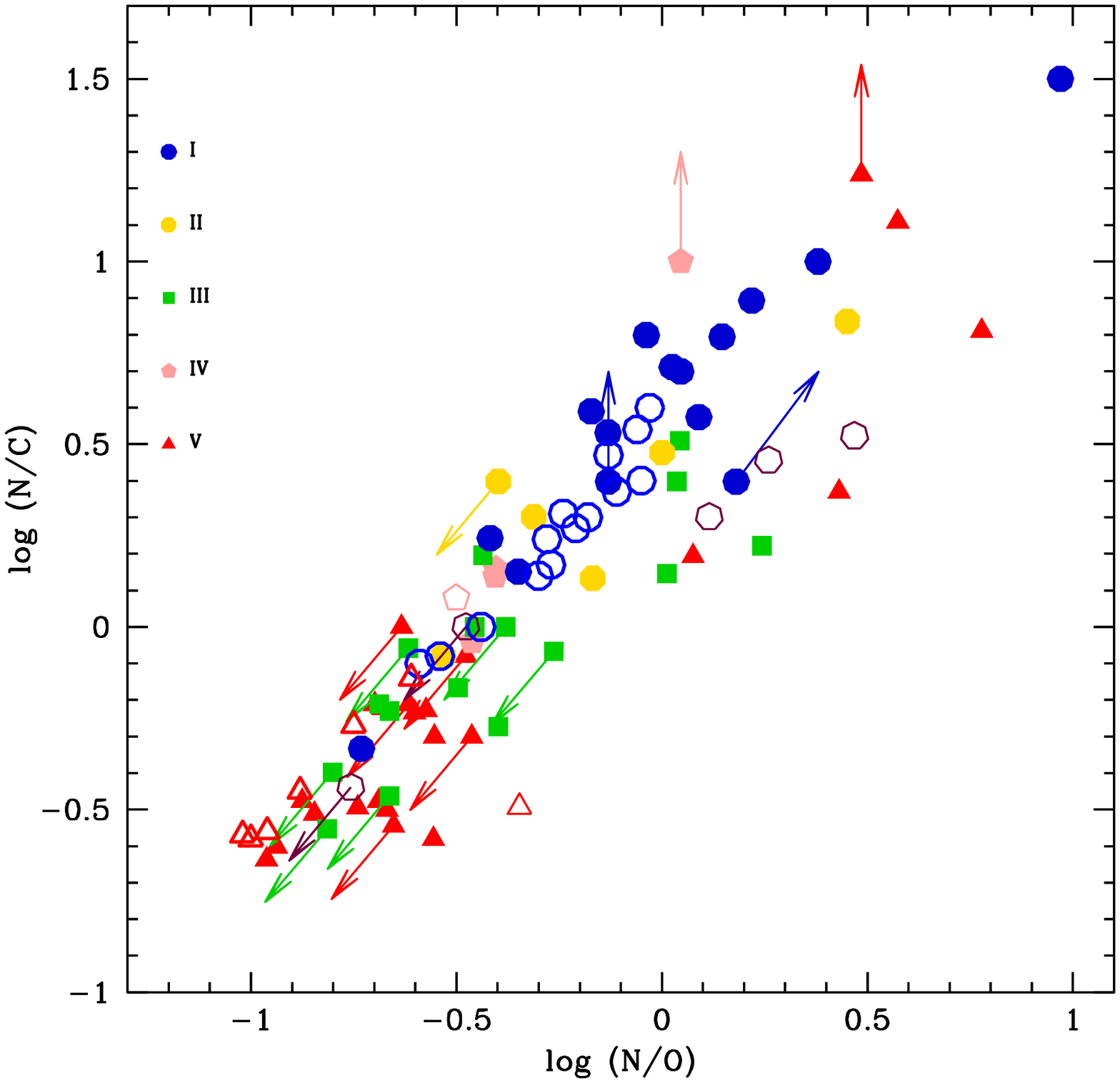}}
     \caption{Comparison between the properties of O dwarfs/supergiants (filled symbols) and the BA dwarfs/supergiants of \citet{przy10} shown by open symbols. Dwarfs are marked by triangles, supergiants by circles. \textit{Left}: \logg\ -- \teff\ diagram. Evolutionary tracks are from \citet{ek12}. The orange rectangle shows the expected position of the BA supergiants when they were just off the main sequence, as the O supergiants of our sample. \textit{Right}: log (N/C) versus log (N/O). The error bars have been omitted for clarity.}
     \label{fig_cno_B}
\end{figure*}

\citet{przy10} performed an abundance determination in a sample of B dwarfs and BA supergiants. They obtained a very tight correlation between N/C and N/O that they interpreted as a clear sign of CNO processing. The slope of the N/C -- N/O relation was found to be in excellent agreement with the theoretical expectation from the case where the CN cycle is at equilibrium. In Fig.\ \ref{fig_cno_B} we compare our results to those of \citet{przy10}. The agreement between both studies is remarkable (see right panel of Fig.\ \ref{fig_cno_B}). The Przybilla et al.\ relation falls perfectly onto the one we obtain for O stars. The dwarfs of Przybilla et al.\ are B-type stars and have thus lower mass than the dwarfs of our sample, as seen in the left panel of Fig.\ \ref{fig_cno_B}. In principle, one expects more mixing in higher mass stars, so the relation log(N/C) versus log (N/O) should be shifted up towards higher nitrogen / lower carbon-oxygen content for the Galactic O dwarfs. We do not observe this trend among the bulk of the stars. Only the early O dwarfs are more chemically evolved, as already noted. This indicates that the effects of initial mass on CNO mixing between $\sim$9 and $\sim$25 \msun\ are not strong, at least in dwarfs. Since these objects are still relatively unevolved, this is not surprising. Examination of the more evolved BA supergiants of \citet{przy10} may be more relevant to check mass effects on chemical mixing. These objects cover a range of initial mass between 7 and 25 \msun\ according to the Geneva tracks (see Fig.\ \ref{fig_cno_B}, left panel). The Galactic O supergiants have initial masses between 25 and 40 \msun. Hence, they represent a group of more massive stars. In Fig.\ \ref{fig_cno_B}, it seems that the BA supergiants are \textit{on average} less chemically evolved than the bulk of the O supergiants. In addition the BA supergiants are more evolved in terms of position along evolutionary tracks (they have lower surface gravities than the O supergiants and are farther away from the end of the main sequence). For a meaningful comparison of the chemical properties with O supergiants, one should consider the BA supergiants when they were just off the main sequence. The rectangle in Fig.\ \ref{fig_cno_B} shows the position of the BA supergiants at that time. There the BA supergiants were most likely less chemically evolved than they are now. Consequently, the possible trend of lower chemical enrichment compared to O supergiant was stronger at that time. All in all, there seems to be a hint that among evolved objects (in terms of distance to the main sequence) more massive stars are more chemically evolved. However, the study a larger sample of OBA supergiants is required to confirm this trend.     

In conclusion, there is good evidence that chemical mixing acts more efficiently at higher masses.

\vspace{0.3cm}

\begin{figure}[]
\centering
\includegraphics[width=9cm]{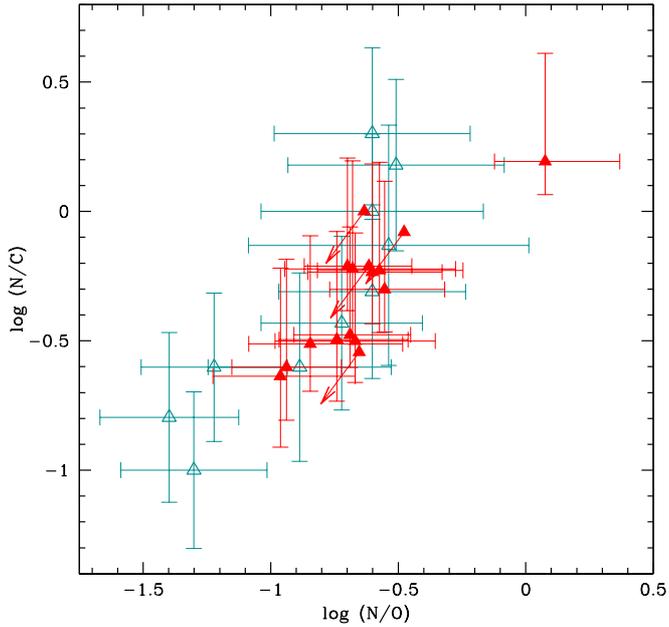}
\caption{log (N/C) as a function of log (N/O) for dwarfs stars with \teff\ $<$ 37500 K. Filled triangles are Galactic stars from our sample; open triangles are SMC dwarfs from \citet{jc13}.}
\label{fig_cno_V}
\end{figure}

\citet{jc13} presented the analysis of the surface abundances of dwarfs in the Small Magellanic Cloud (SMC). Their sample covered spectral types from O4 to O9. Evolutionary calculations predict a stronger enrichment at lower metallicity \citep[e.g.][]{maeder09}. In order to isolate such an effect, we show in Fig.\ \ref{fig_cno_V} the log (N/C) -- log (N/O) diagram for dwarfs in our sample and that of \citet{jc13}. We have restricted ourselves to stars with \teff\ $<$ 37500 K so that all stars have initial masses in the range 18-25 \msun\ (according to the Geneva tracks). Given the error bars, there is no clear trend in this figure: SMC dwarfs do not appear to be significantly more enriched than Galactic stars. Since only a modest enrichment is expected for such objects, this result is not entirely surprising. More evolved giants or supergiants would be better targets for this type of comparison. Unfortunately, there are too few O supergiants at low metallicity for which CNO abundances have been determined.
  
\begin{figure*}[]
     \centering
\subfigure{ \includegraphics[width=.45\textwidth]{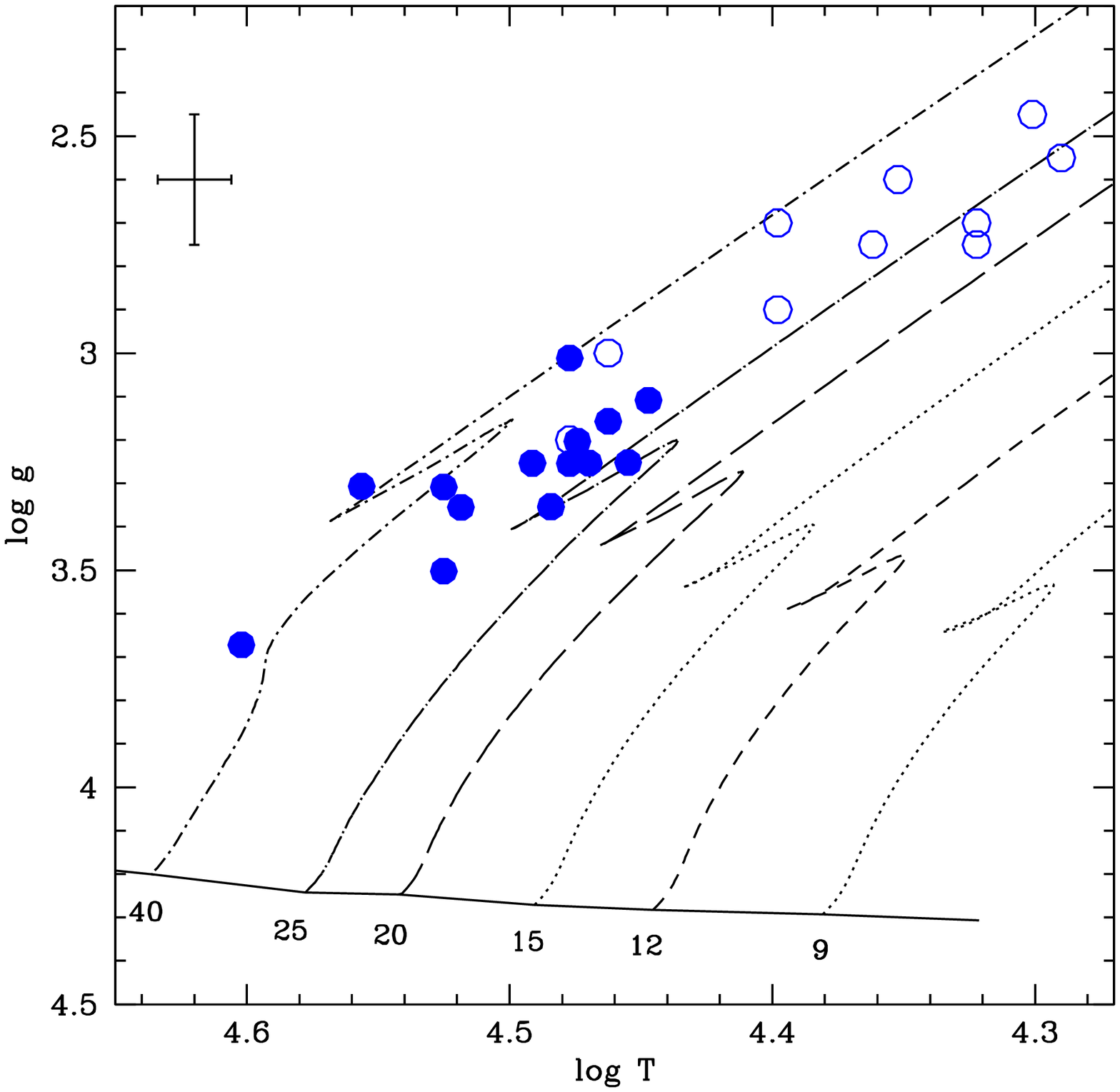}}
     \hspace{0.1cm}
     \subfigure{
           \includegraphics[width=.45\textwidth]{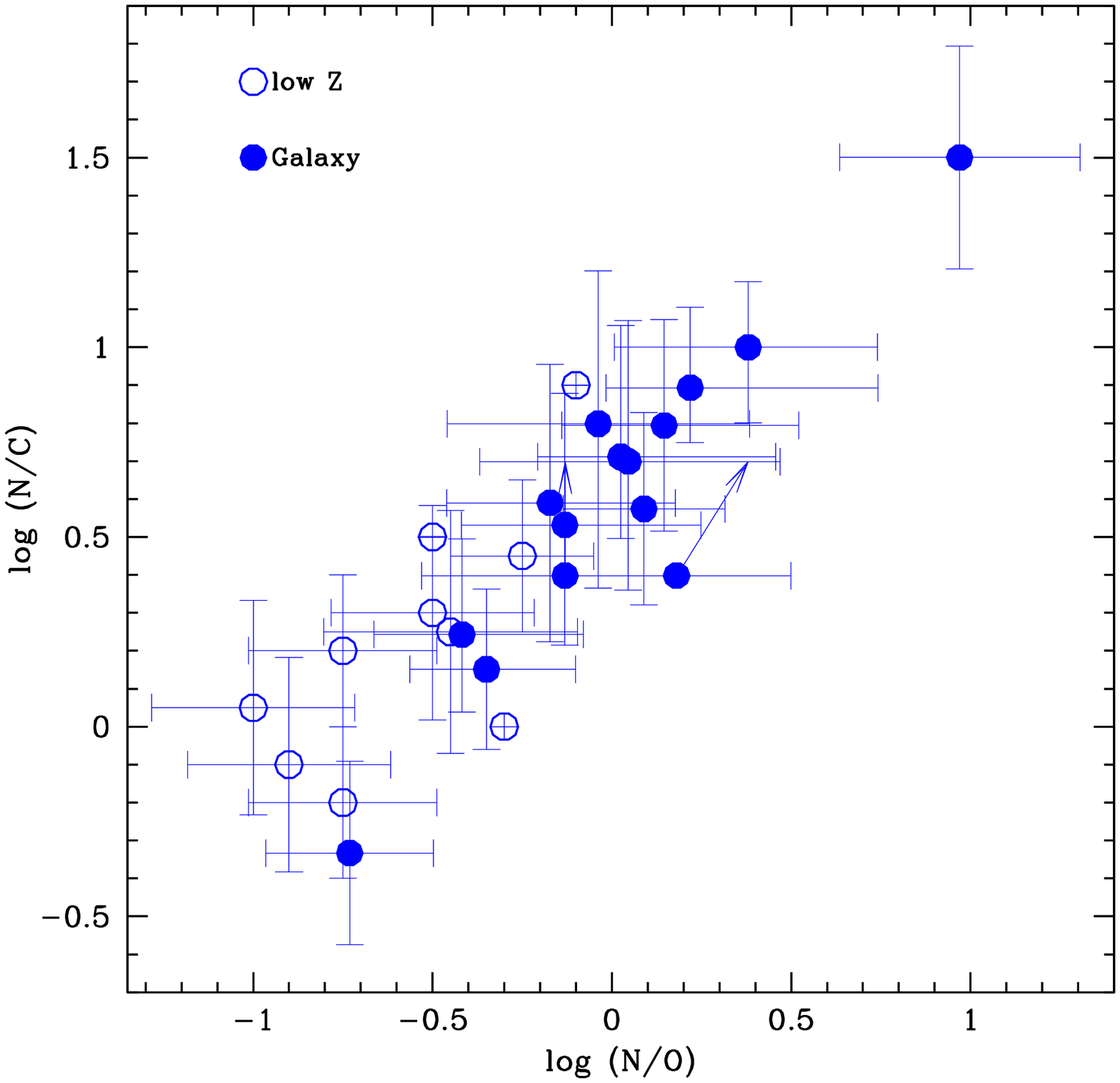}}
     \caption{Comparison between the properties of O supergiants (filled symbols) and the B supergiants at low metallicity of \citet{bre06,bre07} shown by open symbols.  \textit{Left}: \logg\ -- \teff\ diagram. Evolutionary tracks are from \citet{ek12}. \textit{Right}: log (N/C) versus log (N/O).}
     \label{fig_B_lowZ}
\end{figure*}

To further investigate the effects of metallicity on the surface abundances, we consider in Fig.\ \ref{fig_B_lowZ} the O supergiants of our sample together with the B supergiants studied by \citet{bre06,bre07}. These objects are located in external galaxies of the Local Group and are metal-poor, with metallicities $Z \sim 0.1$. In the left panel of Fig.\ \ref{fig_B_lowZ}, these B supergiants have roughly the same range of initial masses as the Galactic O supergiants of the MiMeS sample\footnote{We use evolutionary tracks at solar metallicity, which may not be appropriate for the B supergiants of Bresolin et al. However, the mass range would not differ drastically if we were using low Z tracks.}. Hence, for the same metallicity, one would expect them to be more chemically mixed. In addition, since mixing is predicted to be stronger at low metallicity \citep{mm05}, they should show even higher N/C and N/O ratios. The right panel of Fig.\ \ref{fig_B_lowZ} does not reveal any clear trend. If anything, the low metallicity B supergiants appear somewhat \textit{less} evolved than the O supergiants. Before further interpretation, this results needs to be confirmed by new analysis of O and B supergiants in different environments. We note in particular that the Galactic and low Z stars have not been analyzed homogeneously (we rely on line fitting while Bresolin et al.\ determined abundances from the curve of growth) and systematic differences between methods may exist. 

The present conclusion regarding the effects of metallicity on the strength of mixing is that there does not seem to be any clear trend. To make progress on this topic, samples of O, B and A supergiants with the same initial masses and different metallicities need to be defined and analyzed homogeneously.

%---------------------------------------------------------------
\subsection{Magnetic stars}
\label{s_mag}

We have included seven known magnetic O stars in our abundance study: $\theta^1$Ori~C \citep{donati02,wade06}, HD~57682 \citep{gru09}, HD~108 \citep{martins10},  HD~148937 \citep{hubrig08,wade12a}, HD~191612 \citep{donati06,wade11}, Tr16-22 \citep{naze12} and CPD-28 2561 (Wade et al., submitted). Fig.\ \ref{fig_CN} and \ref{fig_CNO} indicate that these magnetic stars have surface abundances that do not depart from other O-type stars. Fig.\ \ref{fig_CNO} shows that three Of?p stars (HD~108, HD~148937 and HD~191612) are as chemically evolved as O supergiants or some of the most massive O dwarfs. HD~57682 lies in the same region as other subgiants. Only $\theta^1$Ori~C may be shifted towards low values of N/C or high values of N/O, but the error bars do not exclude that it is similar to other O dwarfs. Tr16-22 and CPD-28 2561 have surface abundances consistent with the bulk of O dwarfs. \citet{Omag12} reached similar conclusions based on the nitrogen content of magnetic and comparison stars: magnetic massive stars do not depart from most O stars. Martins et al.\  also pointed out that magnetic O stars rotate on average slowly (perhaps because of magnetic braking). Indeed, the seven O stars listed above have \vsini\ $<$ 40 \kms.

\begin{figure}[]
\centering
\includegraphics[width=9cm]{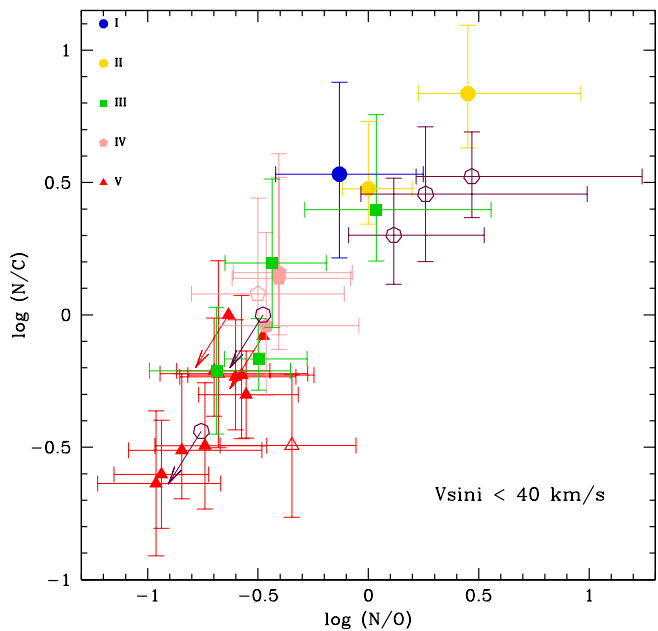}
\caption{log (N/C) as a function of log (N/O) for stars with \vsini\ $<$ 40 \kms. Magnetic stars are shown by open symbols.}
\label{fig_cno_v40}
\end{figure}

In Fig.\ \ref{fig_cno_v40} we show the N/C and N/O ratios for all stars (magnetic and non-magnetic) with projected rotational velocities smaller than 40 \kms. For any magnetic massive star, one can find a non-magnetic star with a similar surface chemical enrichment. The wide range of chemical enrichment covered by magnetic stars is also covered by slowly rotating non-magnetic stars. Consequently, the effect of magnetism on chemical mixing and surface abundances is not clear.

%%%%%%%%%%%%%%%%%%%%%%%%%%%%%%%%%%%%%%%%%%%%%%%%%%%%%%%%%%%%%%%%%%%%%%%%%%%%%%%%%%%%%%%%%%%%%%%%%%%%%%%%%%%%%%%%%%%%%%%%%%%%%%%
%%%%%%%%%%%%%%%%%%%%%%%%%%%%%%%%%%%%%%%%%%%%%%%%%%%%%%%%%%%%%%%%%%%%%%%%%%%%%%%%%%%%%%%%%%%%%%%%%%%%%%%%%%%%%%%%%%%%%%%%%%%%%%%
\section{Conclusions and final remarks}
\label{s_conc}

We have analyzed a sample of 74 Galactic O stars (including seven magnetic objects) observed in the context of the MiMeS survey of massive stars. The sample contains (presumably) single stars, known binaries having been removed. The observations have been performed with the spectropolarimeters ESPaDOnS, NARVAL and HARPSpol respectively at the CFHT, Pic du Midi and La Silla observatory. The spectra cover the optical range from 3800 to at least 7000 \AA. They have signal-to-noise ratios larger than a few hundreds, and a resolution between 65000 and 105000. Using atmosphere models computed with the code CMFGEN we have determined the main surface parameters: effective temperature, surface gravity, rotational velocity, macroturbulent velocity. Our prime focus was to constrain the surface CNO abundances to investigate the chemical evolution of Galactic O stars. Our main results can be summarized as follows:

\begin{itemize}

\item[$\bullet$] there is a clear trend of stronger chemical mixing in more evolved objects. In particular, the ratio of nitrogen to carbon surface abundance is higher in supergiants than in dwarf stars. Giant stars show intermediate degrees of enrichment.  

\item[$\bullet$] the N/C and N/O ratios of Galactic O stars are fully consistent with nucleosynthesis through the CNO cycle. All stars show N/C and N/O ratios intermediate between the limiting cases of partial CN and complete CNO burning.

\item[$\bullet$] among dwarf stars, more massive objects show on average a higher degree of chemical mixing than lower mass objects. This trend is also observed among supergiants when our sample and that of \citet{jc12} are merged. The chemical properties of the B supergiants of \citet{przy10} tend to support this behaviour.

\item[$\bullet$] metallicity effects on the strength of chemical mixing have not been observed when comparing our results to studies conducted in the Magellanic Clouds. This may partially be due to the lack of studies of large samples of evolved O stars at low metallicity. 

\item[$\bullet$] the evolutionary models of \citet{ek12} and \citet{cl13} are equally able to account for the properties of the Galactic O stars. The models of \citet{brott11} predict a chemical enrichment that is too strong in the early phases of evolution.

\item[$\bullet$] 80\% of our sample stars are well accounted for by the predictions of stellar evolution with rotation.

\item[$\bullet$] the effect of magnetism on surface abundances is not clear.

\end{itemize}

Our results show that Galactic O stars have a chemical evolution consistent with CNO nucleosynthesis and chemical mixing (possibly due to rotation). In our study, we have focussed on presumably single O stars to avoid binary effects. The main reason is that predictions of the chemical appearance of binary products are still very uncertain. Very few quantitative predictions have been published so far. The general expectation is that binary interaction will lead to chemical enrichment. \citet{song13} showed that enrichment may appear faster (i.e. at earlier phases, or equivalently at higher \logg) than in single stars. \citet{langer12} indicates that binary models may more easily populate the N/C--N/O diagram than single star models. However, most of the investigations have so far been performed for stars in the mass range 10 to 20 \msun. It may well be that the results can be extrapolated to higher mass stars, but in the absence of quantitative predictions, we cannot perform the same type of tests as those that we performed with models of rotating single stars. 

\citet{dm14} cautioned that many massive stars will experience a merger in an early binary phase, and may thus appear as single stars. If the merger happens sufficiently early, then we do not expect it to affect the chemical properties of the product, since the two  components would be most likely unevolved as our results on O dwarfs indicate. Most of the evolution would then be that of a single star that was formed by a merger. The chemical properties should then result from the physics of single stars. We cannot exclude that some of the stars of our sample are such mergers. \citet{langer12} states (see his Section 3.3) that ``the amount of mixing experienced in massive merger stars is rather unclear''. In absence of quantitative predictions on the appearance of such objects, we cannot test the presence of merger products in our sample. Thus, for the time being, the properties of the Galactic O stars we analyzed appear to be best explained by single star evolution with rotational mixing.

There are good indications that such models are appropriate. The scaling of chemical mixing with stellar mass is one of their predictions that we were able to confirm. However, we stressed that the strongest test, i.e. a direct relation between chemical mixing and rotational velocity, could not be performed in our study (see Sect.\ \ref{s_ab_obs}). This would require a larger number of fast rotating stars and a careful selection of objects with similar masses and age (or luminosity class). Further studies of this type will provide additional tests of rotational mixing among massive stars.

%%#####################################################################
\section*{Acknowledgments}

FM thanks the Agence Nationale de la Recherche for financial support (grant ANR-11-JS56-0007). We thank John Hillier for making his atmosphere code CMFGEN available and for constant help with it. GAW acknowledges Discovery Grant support from the Natural Sciences and Engineering Research Council of Canada (NSERC). We thank an anonymous referee for a prompt and positive report.

%%#####################################################################
\bibliographystyle{aa}
\bibliography{mimes_O}

\begin{thebibliography}{63}
\expandafter\ifx\csname natexlab\endcsname\relax\def\natexlab#1{#1}\fi

\bibitem[{{Aerts} {et~al.}(2014){Aerts}, {Molenberghs}, {Kenward}, \&
  {Neiner}}]{aerts14}
{Aerts}, C., {Molenberghs}, G., {Kenward}, M.~G., \& {Neiner}, C. 2014, \apj,
  781, 88

\bibitem[{{Bouret} {et~al.}(2012){Bouret}, {Hillier}, {Lanz}, \&
  {Fullerton}}]{jc12}
{Bouret}, J.-C., {Hillier}, D.~J., {Lanz}, T., \& {Fullerton}, A.~W. 2012,
  \aap, 544, A67

\bibitem[{{Bouret} {et~al.}(2013){Bouret}, {Lanz}, {Martins}, {Marcolino},
  {Hillier}, {Depagne}, \& {Hubeny}}]{jc13}
{Bouret}, J.-C., {Lanz}, T., {Martins}, F., {et~al.} 2013, \aap, 555, A1

\bibitem[{{Bresolin} {et~al.}(2006){Bresolin}, {Pietrzy{\'n}ski}, {Urbaneja},
  {Gieren}, {Kudritzki}, \& {Venn}}]{bre06}
{Bresolin}, F., {Pietrzy{\'n}ski}, G., {Urbaneja}, M.~A., {et~al.} 2006, \apj,
  648, 1007

\bibitem[{{Bresolin} {et~al.}(2007){Bresolin}, {Urbaneja}, {Gieren},
  {Pietrzy{\'n}ski}, \& {Kudritzki}}]{bre07}
{Bresolin}, F., {Urbaneja}, M.~A., {Gieren}, W., {Pietrzy{\'n}ski}, G., \&
  {Kudritzki}, R.-P. 2007, \apj, 671, 2028

\bibitem[{{Brott} {et~al.}(2011{\natexlab{a}}){Brott}, {de Mink}, {Cantiello},
  {Langer}, {de Koter}, {Evans}, {Hunter}, {Trundle}, \& {Vink}}]{brott11}
{Brott}, I., {de Mink}, S.~E., {Cantiello}, M., {et~al.} 2011{\natexlab{a}},
  \aap, 530, A115

\bibitem[{{Brott} {et~al.}(2011{\natexlab{b}}){Brott}, {Evans}, {Hunter}, {de
  Koter}, {Langer}, {Dufton}, {Cantiello}, {Trundle}, {Lennon}, {de Mink},
  {Yoon}, \& {Anders}}]{brott11b}
{Brott}, I., {Evans}, C.~J., {Hunter}, I., {et~al.} 2011{\natexlab{b}}, \aap,
  530, A116

\bibitem[{{Chieffi} \& {Limongi}(2013)}]{cl13}
{Chieffi}, A. \& {Limongi}, M. 2013, \apj, 764, 21

\bibitem[{{Conti}(1975)}]{conti75}
{Conti}, P.~S. 1975, Memoires of the Societe Royale des Sciences de Liege, 9,
  193

\bibitem[{{Crowther}(2007)}]{paul07}
{Crowther}, P.~A. 2007, \araa, 45, 177

\bibitem[{{Crowther} {et~al.}(2002){Crowther}, {Dessart}, {Hillier}, {Abbott},
  \& {Fullerton}}]{paul02}
{Crowther}, P.~A., {Dessart}, L., {Hillier}, D.~J., {Abbott}, J.~B., \&
  {Fullerton}, A.~W. 2002, \aap, 392, 653

\bibitem[{{de Mink} {et~al.}(2009){de Mink}, {Cantiello}, {Langer}, {Pols},
  {Brott}, \& {Yoon}}]{dm09}
{de Mink}, S.~E., {Cantiello}, M., {Langer}, N., {et~al.} 2009, \aap, 497, 243

\bibitem[{{de Mink} {et~al.}(2014){de Mink}, {Sana}, {Langer}, {Izzard}, \&
  {Schneider}}]{dm14}
{de Mink}, S.~E., {Sana}, H., {Langer}, N., {Izzard}, R.~G., \& {Schneider},
  F.~R.~N. 2014, \apj, 782, 7

\bibitem[{{Donati} {et~al.}(2002){Donati}, {Babel}, {Harries}, {Howarth},
  {Petit}, \& {Semel}}]{donati02}
{Donati}, J.-F., {Babel}, J., {Harries}, T.~J., {et~al.} 2002, \mnras, 333, 55

\bibitem[{{Donati} {et~al.}(2006){Donati}, {Howarth}, {Bouret}, {Petit},
  {Catala}, \& {Landstreet}}]{donati06}
{Donati}, J.-F., {Howarth}, I.~D., {Bouret}, J.-C., {et~al.} 2006, \mnras, 365,
  L6

\bibitem[{{Ekstr{\"o}m} {et~al.}(2012){Ekstr{\"o}m}, {Georgy}, {Eggenberger},
  {Meynet}, {Mowlavi}, {Wyttenbach}, {Granada}, {Decressin}, {Hirschi},
  {Frischknecht}, {Charbonnel}, \& {Maeder}}]{ek12}
{Ekstr{\"o}m}, S., {Georgy}, C., {Eggenberger}, P., {et~al.} 2012, \aap, 537,
  A146

\bibitem[{{Evans} {et~al.}(2006){Evans}, {Lennon}, {Smartt}, \&
  {Trundle}}]{evans06}
{Evans}, C.~J., {Lennon}, D.~J., {Smartt}, S.~J., \& {Trundle}, C. 2006, \aap,
  456, 623

\bibitem[{{Evans} {et~al.}(2005){Evans}, {Smartt}, {Lee}, {Lennon}, {Kaufer},
  {Dufton}, {Trundle}, {Herrero}, {Sim{\'o}n-D{\'{\i}}az}, {de Koter},
  {Hamann}, {Hendry}, {Hunter}, {Irwin}, {Korn}, {Kudritzki}, {Langer},
  {Mokiem}, {Najarro}, {Pauldrach}, {Przybilla}, {Puls}, {Ryans}, {Urbaneja},
  {Venn}, \& {Villamariz}}]{evans05}
{Evans}, C.~J., {Smartt}, S.~J., {Lee}, J.-K., {et~al.} 2005, \aap, 437, 467

\bibitem[{{Gray}(1976)}]{gray76}
{Gray}, D.~F. 1976, {The observation and analysis of stellar photospheres}

\bibitem[{{Grevesse} {et~al.}(2010){Grevesse}, {Asplund}, {Sauval}, \&
  {Scott}}]{ga10}
{Grevesse}, N., {Asplund}, M., {Sauval}, A.~J., \& {Scott}, P. 2010, \apss,
  328, 179

\bibitem[{{Groh} {et~al.}(2013){Groh}, {Meynet}, {Georgy}, \&
  {Ekstr{\"o}m}}]{groh13}
{Groh}, J.~H., {Meynet}, G., {Georgy}, C., \& {Ekstr{\"o}m}, S. 2013, \aap,
  558, A131

\bibitem[{{Grunhut} {et~al.}(2009){Grunhut}, {Wade}, {Marcolino}, {Petit},
  {Henrichs}, {Cohen}, {Alecian}, {Bohlender}, {Bouret}, {Kochukhov}, {Neiner},
  {St-Louis}, \& {Townsend}}]{gru09}
{Grunhut}, J.~H., {Wade}, G.~A., {Marcolino}, W.~L.~F., {et~al.} 2009, \mnras,
  400, L94

\bibitem[{{Heger} {et~al.}(2000){Heger}, {Langer}, \& {Woosley}}]{hl00}
{Heger}, A., {Langer}, N., \& {Woosley}, S.~E. 2000, \apj, 528, 368

\bibitem[{{Hillier} \& {Miller}(1998)}]{hm98}
{Hillier}, D.~J. \& {Miller}, D.~L. 1998, \apj, 496, 407

\bibitem[{{Hubrig} {et~al.}(2008){Hubrig}, {Sch{\"o}ller}, {Schnerr},
  {Gonz{\'a}lez}, {Ignace}, \& {Henrichs}}]{hubrig08}
{Hubrig}, S., {Sch{\"o}ller}, M., {Schnerr}, R.~S., {et~al.} 2008, \aap, 490,
  793

\bibitem[{{Hunter} {et~al.}(2009){Hunter}, {Brott}, {Langer}, {Lennon},
  {Dufton}, {Howarth}, {Ryans}, {Trundle}, {Evans}, {de Koter}, \&
  {Smartt}}]{hunter09}
{Hunter}, I., {Brott}, I., {Langer}, N., {et~al.} 2009, \aap, 496, 841

\bibitem[{{Hunter} {et~al.}(2007){Hunter}, {Dufton}, {Smartt}, {Ryans},
  {Evans}, {Lennon}, {Trundle}, {Hubeny}, \& {Lanz}}]{hunter07}
{Hunter}, I., {Dufton}, P.~L., {Smartt}, S.~J., {et~al.} 2007, \aap, 466, 277

\bibitem[{{Hunter} {et~al.}(2008){Hunter}, {Lennon}, {Dufton}, {Trundle},
  {Sim{\'o}n-D{\'{\i}}az}, {Smartt}, {Ryans}, \& {Evans}}]{hunter08}
{Hunter}, I., {Lennon}, D.~J., {Dufton}, P.~L., {et~al.} 2008, \aap, 479, 541

\bibitem[{{Langer}(2012)}]{langer12}
{Langer}, N. 2012, \araa, 50, 107

\bibitem[{{Maeder} \& {Meynet}(1996)}]{mm96}
{Maeder}, A. \& {Meynet}, G. 1996, \aap, 313, 140

\bibitem[{{Maeder} \& {Meynet}(2000)}]{mm00}
{Maeder}, A. \& {Meynet}, G. 2000, \araa, 38, 143

\bibitem[{{Maeder} \& {Meynet}(2005)}]{mm05b}
{Maeder}, A. \& {Meynet}, G. 2005, \aap, 440, 1041

\bibitem[{{Maeder} {et~al.}(2009){Maeder}, {Meynet}, {Ekstr{\"o}m}, \&
  {Georgy}}]{maeder09}
{Maeder}, A., {Meynet}, G., {Ekstr{\"o}m}, S., \& {Georgy}, C. 2009,
  Communications in Asteroseismology, 158, 72

\bibitem[{{Maeder} {et~al.}(2014){Maeder}, {Przybilla}, {Nieva}, {Georgy},
  {Meynet}, {Ekstr{\"o}m}, \& {Eggenberger}}]{maeder14}
{Maeder}, A., {Przybilla}, N., {Nieva}, M.-F., {et~al.} 2014, \aap, 565, A39

\bibitem[{{Martins} {et~al.}(2010){Martins}, {Donati}, {Marcolino}, {Bouret},
  {Wade}, {Escolano}, {Howarth}, \& {Mimes Collaboration}}]{martins10}
{Martins}, F., {Donati}, J.-F., {Marcolino}, W.~L.~F., {et~al.} 2010, \mnras,
  407, 1423

\bibitem[{{Martins} {et~al.}(2012{\natexlab{a}}){Martins}, {Escolano}, {Wade},
  {Donati}, {Bouret}, \& {Mimes Collaboration}}]{Omag12}
{Martins}, F., {Escolano}, C., {Wade}, G.~A., {et~al.} 2012{\natexlab{a}},
  \aap, 538, A29

\bibitem[{{Martins} \& {Hillier}(2012)}]{mh12}
{Martins}, F. \& {Hillier}, D.~J. 2012, \aap, 545, A95

\bibitem[{{Martins} {et~al.}(2012{\natexlab{b}}){Martins}, {Mahy}, {Hillier},
  \& {Rauw}}]{martins12}
{Martins}, F., {Mahy}, L., {Hillier}, D.~J., \& {Rauw}, G. 2012{\natexlab{b}},
  \aap, 538, A39

\bibitem[{{Martins} \& {Palacios}(2013)}]{mp13}
{Martins}, F. \& {Palacios}, A. 2013, \aap, 560, A16

\bibitem[{{Martins} {et~al.}(2005){Martins}, {Schaerer}, \& {Hillier}}]{msh05}
{Martins}, F., {Schaerer}, D., \& {Hillier}, D.~J. 2005, \aap, 436, 1049

\bibitem[{{Meynet} \& {Maeder}(2005)}]{mm05}
{Meynet}, G. \& {Maeder}, A. 2005, \aap, 429, 581

\bibitem[{{Mokiem} {et~al.}(2007){Mokiem}, {de Koter}, {Evans}, {Puls},
  {Smartt}, {Crowther}, {Herrero}, {Langer}, {Lennon}, {Najarro}, {Villamariz},
  \& {Vink}}]{mokiem07}
{Mokiem}, M.~R., {de Koter}, A., {Evans}, C.~J., {et~al.} 2007, \aap, 465, 1003

\bibitem[{{Mokiem} {et~al.}(2006){Mokiem}, {de Koter}, {Evans}, {Puls},
  {Smartt}, {Crowther}, {Herrero}, {Langer}, {Lennon}, {Najarro}, {Villamariz},
  \& {Yoon}}]{mokiem06}
{Mokiem}, M.~R., {de Koter}, A., {Evans}, C.~J., {et~al.} 2006, \aap, 456, 1131

\bibitem[{{Naz{\'e}} {et~al.}(2012){Naz{\'e}}, {Bagnulo}, {Petit}, {Rivinius},
  {Wade}, {Rauw}, \& {Gagn{\'e}}}]{naze12}
{Naz{\'e}}, Y., {Bagnulo}, S., {Petit}, V., {et~al.} 2012, \mnras, 423, 3413

\bibitem[{{Nieva} \& {Przybilla}(2006)}]{np06}
{Nieva}, M.~F. \& {Przybilla}, N. 2006, \apjl, 639, L39

\bibitem[{{Nieva} \& {Przybilla}(2007)}]{np07}
{Nieva}, M.~F. \& {Przybilla}, N. 2007, \aap, 467, 295

\bibitem[{{Nieva} \& {Przybilla}(2014)}]{nieva14}
{Nieva}, M.-F. \& {Przybilla}, N. 2014, \aap, 566, A7

\bibitem[{{Petrovic} {et~al.}(2005){Petrovic}, {Langer}, \& {van der
  Hucht}}]{petro05}
{Petrovic}, J., {Langer}, N., \& {van der Hucht}, K.~A. 2005, \aap, 435, 1013

\bibitem[{{Przybilla} {et~al.}(2006){Przybilla}, {Butler}, {Becker}, \&
  {Kudritzki}}]{przy06}
{Przybilla}, N., {Butler}, K., {Becker}, S.~R., \& {Kudritzki}, R.~P. 2006,
  \aap, 445, 1099

\bibitem[{{Przybilla} {et~al.}(2010){Przybilla}, {Firnstein}, {Nieva},
  {Meynet}, \& {Maeder}}]{przy10}
{Przybilla}, N., {Firnstein}, M., {Nieva}, M.~F., {Meynet}, G., \& {Maeder}, A.
  2010, \aap, 517, A38

\bibitem[{{Rivero Gonz{\'a}lez} {et~al.}(2011){Rivero Gonz{\'a}lez}, {Puls}, \&
  {Najarro}}]{rg11}
{Rivero Gonz{\'a}lez}, J.~G., {Puls}, J., \& {Najarro}, F. 2011, \aap, 536, A58

\bibitem[{{Rivero Gonz{\'a}lez} {et~al.}(2012){Rivero Gonz{\'a}lez}, {Puls},
  {Najarro}, \& {Brott}}]{rg12}
{Rivero Gonz{\'a}lez}, J.~G., {Puls}, J., {Najarro}, F., \& {Brott}, I. 2012,
  \aap, 537, A79

\bibitem[{{Sim{\'o}n-D{\'{\i}}az} \& {Herrero}(2007)}]{sergio07}
{Sim{\'o}n-D{\'{\i}}az}, S. \& {Herrero}, A. 2007, \aap, 468, 1063

\bibitem[{{Smith} \& {Owocki}(2006)}]{smith06}
{Smith}, N. \& {Owocki}, S.~P. 2006, \apjl, 645, L45

\bibitem[{{Song} {et~al.}(2013){Song}, {Maeder}, {Meynet}, {Huang},
  {Ekstr{\"o}m}, \& {Granada}}]{song13}
{Song}, H.~F., {Maeder}, A., {Meynet}, G., {et~al.} 2013, \aap, 556, A100

\bibitem[{{Sota} {et~al.}(2014){Sota}, {Ma{\'{\i}}z Apell{\'a}niz}, {Morrell},
  {Barb{\'a}}, {Walborn}, {Gamen}, {Arias}, \& {Alfaro}}]{sota14}
{Sota}, A., {Ma{\'{\i}}z Apell{\'a}niz}, J., {Morrell}, N.~I., {et~al.} 2014,
  \apjs, 211, 10

\bibitem[{{Sota} {et~al.}(2011){Sota}, {Ma{\'{\i}}z Apell{\'a}niz}, {Walborn},
  {Alfaro}, {Barb{\'a}}, {Morrell}, {Gamen}, \& {Arias}}]{sota11}
{Sota}, A., {Ma{\'{\i}}z Apell{\'a}niz}, J., {Walborn}, N.~R., {et~al.} 2011,
  \apjs, 193, 24

\bibitem[{{Tramper} {et~al.}(2013){Tramper}, {Gr{\"a}fener}, {Hartoog}, {Sana},
  {de Koter}, {Vink}, {Ellerbroek}, {Langer}, {Garcia}, {Kaper}, \& {de
  Mink}}]{tramper13}
{Tramper}, F., {Gr{\"a}fener}, G., {Hartoog}, O.~E., {et~al.} 2013, \aap, 559,
  A72

\bibitem[{{Trundle} {et~al.}(2007){Trundle}, {Dufton}, {Hunter}, {Evans},
  {Lennon}, {Smartt}, \& {Ryans}}]{trundle07}
{Trundle}, C., {Dufton}, P.~L., {Hunter}, I., {et~al.} 2007, \aap, 471, 625

\bibitem[{{Vink} {et~al.}(2001){Vink}, {de Koter}, \& {Lamers}}]{vink01}
{Vink}, J.~S., {de Koter}, A., \& {Lamers}, H.~J.~G.~L.~M. 2001, \aap, 369, 574

\bibitem[{{Wade} {et~al.}(2006){Wade}, {Fullerton}, {Donati}, {Landstreet},
  {Petit}, \& {Strasser}}]{wade06}
{Wade}, G.~A., {Fullerton}, A.~W., {Donati}, J.-F., {et~al.} 2006, \aap, 451,
  195

\bibitem[{{Wade} {et~al.}(2012){Wade}, {Grunhut}, {Gr{\"a}fener}, {Howarth},
  {Martins}, {Petit}, {Vink}, {Bagnulo}, {Folsom}, {Naz{\'e}}, {Walborn},
  {Townsend}, \& {Evans}}]{wade12a}
{Wade}, G.~A., {Grunhut}, J., {Gr{\"a}fener}, G., {et~al.} 2012, \mnras, 419,
  2459

\bibitem[{{Wade} {et~al.}(2011){Wade}, {Howarth}, {Townsend}, {Grunhut},
  {Shultz}, {Bouret}, {Fullerton}, {Marcolino}, {Martins}, {Naz{\'e}}, {Ud
  Doula}, {Walborn}, \& {Donati}}]{wade11}
{Wade}, G.~A., {Howarth}, I.~D., {Townsend}, R.~H.~D., {et~al.} 2011, \mnras,
  416, 3160

\end{thebibliography}
%%#####################################################################
%%#####################################################################

\end{document}